\documentclass[a4paper,12pt]{article}
\usepackage{jheppub}
\usepackage{amsmath,amssymb,latexsym}
\usepackage{url}
\begin{document}
\title{A membrane paradigm at large D}
\author[a]{Sayantani Bhattacharyya,} 
\author[b] {Anandita De,}
\author[c]{Shiraz Minwalla,}
\author[c]{Ravi Mohan}
\author[c]{and Arunabha Saha}
\affiliation[a]{Indian Institute of Technology Kanpur, Kanpur, India-208016}
\affiliation[b]{Indian Institute of Science Education and Research Pune, Pune, India-411008}
\affiliation[c]{Tata Institute of Fundamental Research, Mumbai, India-400005}
\emailAdd{ sayanta@iitk.ac.in}
\emailAdd{ananditade@students.iiserpune.ac.in}
\emailAdd{minwalla@theory.tifr.res.in}
\emailAdd{ravimohan1991@gmail.com}
\emailAdd{arunabha@theory.tifr.res.in}

\abstract{We study $SO(d+1)$ invariant solutions of the classical vacuum Einstein 
	equations in $p+d+3$ dimensions. 
	In the limit $d \to \infty$ with $p$ held fixed we construct a class of 
	solutions labelled by the shape of a membrane (the event horizon), 
	together with a `velocity' field that lives on this membrane. 
	We demonstrate that our metrics can be corrected to nonsingular solutions 
	at first sub-leading order in $\frac{1}{d}$ if and only 
	if the membrane shape and `velocity' field obey equations of motion which 
	we determine. These equations define a well posed initial value 
	problem for the membrane shape and this `velocity' and so completely determine
	the dynamics of the black hole. They may be viewed as governing the 
        non-linear 
        dynamics of the light quasi normal modes of Emparan, Suzuki and Tanabe.}

\maketitle

\section{Introduction}

The rich classical dynamics of uncharged black holes is governed by the 
vacuum Einstein equation
\begin{equation}\label{ev}
R_{\mu\nu}=0.
\end{equation}
This innocuous looking equation captures very  complicated
processes; for example the collision of two black holes and their subsequent 
merger accompanied by gravitational radiation. We owe much of our 
current understanding of these processes to numerics. However the numerics
involved are very challenging; their complexity makes it 
impractical to densely fill the space of initial conditions with numerical 
solutions. It would thus be useful to have analytic techniques to analyse
these solutions. However the phenomena described are of such complexity that 
it seems unlikely that exact analytic solutions will ever be obtained. 
Moreover perturbation theory appears to be ruled out by the fact that Einstein's equations lack a parameter. 

In this paper we follow the lead of  Emparan, Suzuki and Tanabe (EST) 
to introduce a parameter into Einstein's equations. We do this by studying 
these equations in $D$ dimensions. The equations of black hole dynamics 
simplify in this limit and admit a systematic expansion in $\frac{1}{D}$. 
Our general strategy is similar to that of t'Hooft \cite{'tHooft:1973jz} 
who introduced a parameter into the study of Yang Mills theory by 
replacing the $SU(3)$ gauge group by $SU(N)$, and that of Witten who 
introduced a parameter into the study of atomic and molecular physics by 
analysing the quantum mechanics of the $\frac{1}{r}$ potential in large 
dimensions \cite{Witten:1980pt}.

EST and collaborators  have recently pointed out \cite{Emparan:2013moa,Emparan:2013xia,Emparan:2013oza,Emparan:2014cia,Emparan:2014jca,Emparan:2014aba,Emparan:2015rva,Emparan:2015eur} 
that several features of black hole dynamics simplify in the limit of a large 
number of spacetime dimensions $D$. A Schwarzschild black hole with 
Schwarzschild radius $r_0$ in $D$ 
spacetime dimensions is described by the metric 
\begin{equation}\label{schbh}
ds^2= -\left( 1-\left( \frac{r_0}{r} \right)^{D-3} \right) dt^2
+ \frac{dr^2}{\left( 1-\left( \frac{r_0}{r} \right)^{D-3} \right)}
+ r^2 d\Omega_{D-2}^2.
\end{equation}
If $r$ is held fixed at a value greater than $r_0$ as $D \to \infty$ then  
$\left( \frac{r_0}{r} \right)^{D-3} \to 0$ and the metric \eqref{schbh}
reduces to flat space. On the other hand if we set $r=r_0(1+\frac{R}{D-3} )$
and hold $R$ fixed as $D$ is taken to infinity then  
$\left( \frac{r_0}{r} \right)^{D-3}  \to e^{-R}$. It follows that the 
gravitational tail of a black hole in $D$ spacetime dimensions extends only 
over a distance $\frac{r_0}{D-3}$ away from its event horizon \cite{Emparan:2013moa}, 
a distance we will refer to as the `thickness of the membrane'. 

EST have computed spectrum of quasinormal modes
of the black hole \eqref{schbh} in an expansion in $\frac{1}{D}$
\cite{Emparan:2014aba,Emparan:2014cia}. 
They find that almost all of the infinite number of  
quasinormal modes at every angular momentum 
have frequencies of order the inverse of the membrane thickness, i.e. of 
order $\frac{D}{r_0}$. However at each angular momentum a small number of 
modes (two scalars and one vector at each value of the angular momentum) are 
much lighter than the generic mode; the frequencies of the light modes 
are of order $\frac{1}{r_0}$. Moreover the light quasinormal modes are 
supported entirely inside the membrane region.
\footnote{In contrast the heavy quasinonrmal modes are non-trivial outside 
the membrane region where they reduce to purely outgoing modes. }

The separation in scale between the heavy and light quasinormal modes suggests
that the full nonlinear dynamics of a black hole over time scales much larger 
than $\frac{r_0}{D}$ is governed by an effective non-linear theory of the 
light quasinormal modes. We expect the degrees of freedom of this theory 
to reside within the membrane region, and so effectively on a
codimension one surface in spacetime at length scales large compared to 
$\frac{r_0}{D}$.

In this paper we will verify the expectation of the previous
paragraph by explicit analysis of Einstein's equations. We identify the
auxiliary dynamical system that lives on the membrane and determine its 
dynamical equations of motion. Our work has several similarities to the 
Fluid-Gravity correspondence (\cite{Bhattacharyya:2008jc} see 
\cite{Rangamani:2009xk,Hubeny:2011hd} for reviews), however there are also 
significant differences. The construction of this paper applies to gravity even in flat space. The effective `hydrodynamical' theory obtained in this paper 
lives on a fluctuating  surface rather than the fixed field theory background 
of fluid gravity. Finally  the constructions presented in this paper are \cite{Armas:2013goa}
justified {\it not} by an expansion in gradients in units of the horizon 
radius, but rather by an expansion in $\frac{1}{D}$. Our membranes also 
has several conceptual similarities to the blackfolds (see e.g. 
\cite{Emparan:2011hg},\cite{Camps:2012hw}, \cite{Armas:2013goa}, \cite{Armas:2013hsa}), but differ from the later in decoupling from radiation.

A solution of Einstein's equations in $D$ dimensions is given 
by a metric tensor with $D(D+1)/2$ components; each of these components is 
a function of $D$ spacetime variables. It is clear that the generic solution 
of this sort has no analogue at any fixed finite value of $D$. 
For example a metric in four dimensions has many fewer legs and and 
many fewer directions in which to wiggle than a metric in a million dimensions.
In order to take a sensible large $D$ limit we write $D=p+d+3$ and 
divide up the $D$ spacetime dimensions into two groups. 
The first group contains $p+2$ dimensions including time and 
the second contains the remaining $d+1$ dimensions. We restrict our 
attention {\it only} to solutions that preserve the $SO(d+1)$ rotational 
symmetry in $d+1$ dimensions. We then take the limit $d \to \infty$ 
with $p$ held fixed.  \footnote{It follows that the solutions we study 
have analogues in all dimensions greater than or equal to $p+3$. }

In the rest of this paper we use the following notation. Let $w_a$ 
represent the first set of $p+2$ coordinates. Let $z_M$ denote the last 
set of $d+1$ coordinates. The metric of flat $D$ dimensional spacetime is  
\begin{equation}\label{fss} \begin{split}
ds^2&=d w_a dw^a + dz_M dz^M=dw_a dw^a + dS^2+S^2 d\Omega_{d}^2\\
S^2&=z_M z^M
\end{split}
\end{equation}
Let $x^\mu=(S, w^a)$ so that the index $\mu$ runs over $p+3$ variables. 
The most general metric that preserves the $SO(d+1)$ isometry takes the form
\begin{equation}\label{redn}
ds^2_{full} = g_{\mu\nu} dx^\mu dx^\nu + e^\phi d \Omega_{d}^2
\end{equation}
where $g_{\mu\nu}$ and $\phi$ respectively are  a metric and 
a scalar field in the $p+3$ dimensional space spanned by $x^\mu$; 
in particular they are functions only of $x^\mu$ and not of the 
coordinates of the $d$ sphere. 

Below we derive the equations of motion 
for $g_{\mu\nu}$ and $\phi$, and explain that these equations admit an 
interesting large $d$ limit. We then explicitly construct a large family of 
solutions of these equations. The solutions we construct are in one to one 
correspondence with the solutions of an auxiliary problem of membrane dynamics 
in flat space that we now describe. 

Our auxiliary dynamical system lives on a codimension one 
membrane that resides  in the flat space \eqref{fss}. 
This membrane is free to fluctuate, subject to 
the requirement that it preserves $SO(d+1)$ invariance. The
 position of the membrane can always be characterized by the zeroes of a 
function $B(x^\mu)$ and $SO(d+1)$ invariance forces $B(x^\mu)$ to be a function 
only of $w^a$ and $S$ (and not of the angular coordinates on the $d$ 
sphere). We could, for example, choose the function $B$ to take 
the form $B=S-g(w^a)$ so that the membrane surface is given simply 
by the equation $S=g(w^a)$. It follows that the data in the shape of the 
membrane surface is contained in a single function of $p+2$ variables.
  
For many purposes we can think of the membrane as propagating in the 
auxiliary flat space \footnote{However the $S$ coordinate is inequivalent to the other 
$w^a$ 
coordinates and so our effective dynamical system enjoys invariance only 
	under the  $p+2$  dimensional Lorentz invariance.}.
\begin{equation}\label{auxsp}
ds^2= dw_a dw^a + d S^2
\end{equation}
 Let $n_\mu$ denote 
the outward pointing unit normalized normal one form  field of the membrane. 
$n_\mu$ is assumed always to be spacelike. $n_\mu$ is of course completely 
determined by the shape of the membrane surface; in terms of the function 
$B$ we have $n_\mu= \frac{dB}{\sqrt{dB.d B}}$. 

Our membrane is also equipped with a `velocity' field $u_\mu$ on its world volume
\footnote{More precisely our membrane has a null oneform field $O$ living on its world 
volume. It turns out that the overall scale of $O$ can be absorbed into a subleading shift of the 
membrane surface and so is irrelevant at leading order. Upto a scaling this oneform 
can be cast in the form $O= dS- u_a dw^a$; the fact that $O.O=0$ then implies that $u.u=-1$.
This normalization leads us to refer to the collective coordinate $u_\mu$ as a `velocity'.
The precise physical significance of $u_\mu$ will be clear once we understand its contribution 
to the membrane stress tensor. We leave this to future work. } 
 $u_\mu$ is a $p+3$ dimensional one form field that is further constrained by 
the following unusual conditions
\begin{equation} \label{velconst} \begin{split}
&u^2=-1\\
&u.n= n_S-\frac{1}{n_S}\\
&u_S=0\\
\end{split}
\end{equation}
(here $n_S$ and $u_S$ are simply the $S$ components of the one forms $n$ and $u$). 
The first of \eqref{velconst} simply asserts that $u_\mu$ is a timelike one form
of unit norm, as is usually true of a velocity field. The second of 
\eqref{velconst} asserts that the normal component of $u_\mu$ is not 
free but is determined by the other components of $u_\mu$ and the 
angle the membrane normal vector makes with a unit vector in the $S$ direction. 
The third of \eqref{velconst} asserts that the velocity 
field has no component in the special $S$ direction. The velocity 
one form field $u_\mu$ is specified by $p+3$ functions that live on the membrane.
These $p+3$ equations are constrained by 
by the 3 equations \eqref{velconst} so that the data in the velocity is 
effectively $p$ functions of $p+2$ membrane coordinates. 

The shape of the membrane and the velocity field described above are the 
variables of our auxiliary dynamical system. As we will demonstrate below, 
the equations of this system turn out to be
\begin{equation}\begin{split}\label{mainequations}
& U_{\perp} . K. U_{\perp}   + n_S(n_S^2-1)/S =0 \\
& {\cal P}_a^b \left(  U_\perp . \nabla u_b \right) = 0
\end{split}
\end{equation}
where 
\begin{equation}\label{Vdef} 
U_\perp= U- (U.n) n, ~~~U= dS+ n_S^2(dS-u_\mu dx^\mu)
\end{equation}
$K_{\mu\nu}$ is the extrinsic curvature of the surface, the symbol 
dot ($.$) denotes 
contraction of indices (all indices are raised and lowered by the 
flat metric \eqref{fss} )  and 
${\cal P}_a^b$ is the projector orthogonal to the $3$ dimensional subspace 
spanned by the three one forms $u_\mu$, $n_\mu$ and $dS$ (so that 
${\cal P}_a^b$ projects onto a $p$ dimensional subspace). 
\eqref{mainequations} are 
a set of $p+1$ equations that define an initial value problem for 
$p+1$ variables (the shape of the surface and the constrained 
one form field $u_\mu$). These equations 
are of first order in time derivatives of velocity but second order in 
time derivatives of the shape function, and so the data that determines a 
solution of these equations is initial shape of the membrane and its first time derivative 
together with the value of the velocity field at any instant of time.

The equations (\ref{mainequations}) are the main result of this paper. These equations are completely new and capture the nonlinear dynamics of the slow quasi normal modes of EST and collaborators (which previously had only been understood at the linearized level). We have learnt that the authors of the upcoming paper \cite{Emparan:2015mem} have independently obtained a set of nonlinear equations that describe static large d membranes. While the  membrane equations \eqref{mainequations} of this paper apply to aribitrary dynamical configurations, it should be possible to specialize these equations and thereby rederive  the results of \cite{Emparan:2015mem} (at leading order in $1/d$)  as a special case; we leave this exercise to future work. 

It has, of course, been previously noticed that the dynamics of black hole
horizons resemble the equations of hydrodynamics on a membrane. The fact that the dynamics 
of black hole horizons is governed at large $d$ by the precise 
equations \eqref{mainequations} can be regarded as a mathematical realization 
of this idea in an appropriate context. In the title of this paper we have used the phrase
`membrane paradigm' to summarize our conclusion that black hole dynamics is precisely 
captured by an effective theory on a membrane. It would be interesting to 
investigate the relationship between the equations \eqref{mainequations} 
and the work initiated in \cite{Damour:1978cg,Price:1988sci} (see also 
\cite{Bredberg:2010ky} and subsequent work) that is more 
conventionally referred to as the membrane paradigm.

As a test of our equations we have computed the shape function, normal 
vector fields, extrinsic curvatures and effective velocity one form fields 
for Schwarzschild black holes as well as Myers-Perry black holes (with 
rotations in upto $[\frac{p+1}{2}]$ planes) at large $D$, 
and have explicitly verified that they 
obey the equations \eqref{mainequations}. We have also linearized 
\eqref{mainequations} about the Schwarzschild solutions and computed the 
spectrum of small fluctuations; our results perfectly match the spectrum 
of quasinormal modes computed in \cite{Emparan:2014aba}. 
\footnote{Note that the quasinormal frequencies have 
imaginary pieces at the leading nontrivial order in the $\frac{1}{d}$ 
expansion. It follows that the equation \eqref{mainequations} are inherently 
dissipative, in contrast to  the equations of hydrodynamics 
which are non dissipative at leading (perfect fluid) order.}
 
The reader might wonder how it is consistent to have a membrane fluctuating 
around in a theory of gravity without emitting gravitational radiation. 
We believe that the answer to this question is that our membrane does 
emit radiation, but that the amplitude of the radiation is of order 
$e^{-d}$ any finite distance away from the membrane.  
The radiation may presumably be regarded as resulting from an effective 
coupling of the form $$ \int d^{p+3} x T_{\mu\nu}h^{\mu\nu}$$
Here $h$ is the metric fluctuation to the exterior of the membrane.
$T_{\mu\nu}$ is the effective stress tensor of the membrane 
which is delta function localized on the membrane surface and must be 
conserved as a consequence of  \eqref{mainequations}. It would be very 
interesting to derive this structure (and in particular the formula 
for $T_{\mu\nu}$) in detail. We postpone this task to future work. 

The discussion of the previous paragraph would appear to suggest that 
our membrane loses very little energy to radiation. While this is 
possible it also possible 
that the smallness of the radiation field away from the membrane is actually
a consequence of the extremely rapid fall off of a finite energy radiation 
field due to the large $d$ version of the the inverse 
square law. Restated, the highly dissipative nature of our membrane dynamics 
may be  entirely due to ``radiation'' into the horizon (as in the case of 
the fluid-gravity correspondence) but may also be partly due to radiation 
to infinity. We postpone further discussion of this important 
issue to future work. 

The equations \eqref{mainequations} are the principal results of this paper.
Our equations suggest many interesting questions and generalizations. 
If the loss of energy to infinity is small, it may be possible to recast 
\eqref{mainequations} as the 
equations of conservation of the membrane stress tensor; this would yield 
a formulation of membrane dynamics analogous to the usual formulation of 
hydrodynamics. It should be possible to imitate the analysis of 
\cite{Bhattacharyya:2008xc} to uplift the area form on the event horizon to 
an entropy current on the membrane whose divergence is point wise positive 
for every consistent solution to the 
equations of motion. It would be interesting to find the first correction 
in $\frac{1}{d}$ to \eqref{mainequations} and also to find the analogue 
of \eqref{mainequations} in the Einstein-Maxwell system. Finally it would
be interesting to study the phenomenology of these equations; for instance
to use them to study the head on collision of two spherical black holes.
We leave all these questions to future work.

\section{Membrane like solutions in the large $d$ limit}

\subsection{Dimensional reduction of $SO(d+1)$ invariant metrics}

As we have explained in the introduction, in this paper we study the vacuum 
Einstein equations \eqref{ev},  restricting  attention to solutions that 
preserve $SO(d+1)$ invariance, i.e solutions of the form \eqref{redn}. 
The effective action for the metric $g_{\mu\nu}$ and the scalar field 
$\phi$ in \eqref{redn} takes the form 
\begin{equation}\label{redlagrangian}
 S= \int \sqrt{g} e^{\frac{d \phi}{2}} \left( R + d(d-1) e^{-\phi} + \frac{d(d-1)}{4} (\partial \phi)^2 \right)
\end{equation}
The equations of motion that follow from the variation of this 
action may be shown to take the form (see Appendix A) 
\begin{equation}\label{redeom}\begin{split}
& e^{-\phi}(d-1) - \frac{d}{4} (\partial \phi)^2 - \frac{1}{2}\nabla^2 \phi
 =0\\
&R_{\mu\nu}= \frac{d}{2} \nabla_\mu \nabla_\nu \phi +\frac{d}{4} \nabla_\mu \phi \nabla_\nu \phi  \\
\end{split}
\end{equation}

\subsection{A large $d$ limit} \label{lld}

We are interested in \eqref{redeom} in the large $d$ limit. A glance 
at \eqref{redeom} shows that $\phi$ is not on the same footing as $g_{\mu\nu}$ in this limit; derivatives of $\phi$ are weighted with 
an additional factor of $d$ compared to derivatives of the metric 
$g_{\mu\nu}$. \footnote{Recall that 
 $d \nabla_\mu \nabla_\nu \phi$ has a term involving one derivative and 
one Christoffel symbol, and so should be thought of as having one $\phi$ 
derivative and one metric derivative at leading order.} 
\footnote{The intuitive reason for this difference is clear; 
$\phi$ controls the size of a $d$ sphere. In the large $d$ limit we should 
thus expect derivatives of $\phi$ to be more `expensive' than those 
of the metric.} 
It follows that nontrivial solutions in the  large $d$ limit must involve 
two length scales; a short length scale of order $\frac{1}{d}$ for the 
variations of the metric $g_{\mu\nu}$ and a longer length scale of order 
unity for the variations of $\phi$. \footnote{ 
$\frac{1}{\sqrt{ (\nabla \phi)^2}}$ gives a covariant estimate for the longer
length scale; the square of the Riemann tensor gives a covariant estimate 
for the inverse fourth power
of the shorter length scale.} It is thus possible - and very natural - to choose $x^\mu$ coordinates for the equations  in which the metric and $\phi$ are both of order unity, but derivatives of $\phi$  are 
of order unity while those of the metric are of order $d$. We choose to 
employ such a coordinate system in what follows. 

How can we describe a metric which varies on the length scale $\frac{1}{d}$ in the large $d$ 
limit, and yet is nontrivial over macroscopic length scales?
We employ the following strategy. Choose a particular point $x_0^\mu$ and 
blow up a region of size $\frac{1}{d}$ about this point to unit size
and then appropriately rescale the metric and dilaton gradient. In equations
\begin{equation}\label{cc} \begin{split}
x^\mu&=x_0^\mu + \alpha_a^\mu \frac{y^a}{d}\\
G_{ab}& \equiv d^2 g_{ab}; ~~~g_{\mu\nu}= d^2 \alpha_\mu^a \alpha_\nu^b g_{ab}= 
\alpha_\mu^a \alpha_\nu^b G_{ab} \\
\chi_a& \equiv \nabla_a \phi \times d = \alpha_a^\mu \nabla_\mu \phi
\end{split}
\end{equation}
\footnote{We rescale $\nabla \phi$ in order to account for the extra factor of 
$d$ in derivatives of $\phi$ as compared to derivatives of the metric.}
We must then repeat the procedure of this paragraph for many 
$x_0^\mu$ (so that the union of the patches about each of these points 
is the full manifold) and make sure that descriptions in distinct 
patches agree where they overlap.

Focussing on the patch around $x_0^\mu$, \eqref{redeom} may be rewritten 
as 
\begin{equation}\label{rrreom}\begin{split}
{\frac{1}{2}}& \nabla_a \chi^a=e^{-\phi} \frac{d-1}{d} - \frac{1}{4} \chi^2 \\
&R_{ab}= \frac{1}{2} \nabla_a \chi_b +\frac{1}{4 d} \chi_a \chi_b\\
\end{split}
\end{equation}
where \eqref{rrreom} is written regarding $G_{ab}$ as the metric (curvatures, 
Christoffel symbols and covariant derivatives are also constructed out of
 $G_{ab}$.) 

At leading order \eqref{rrreom} reduces to 
\begin{equation}\label{ldeom}\begin{split}
& -2 G^{ab} \Gamma_{ab}^c\chi_c 
= 4 e^{-\phi}  - \chi^2 \\
&2 R_{ab}= -\Gamma_{ab}^c \chi_c \\
\end{split}
\end{equation}
\footnote{Note that 
$$ -\Gamma^b_{cd} G^{cd} \equiv 
\frac{1}{\sqrt{G}} \partial_{a} \left( G^{ab} \sqrt{G} \right). $$}
where $\phi$ and $\chi_a$ are constants. Note that $d$ has disappeared 
from the leading order equations \eqref{ldeom}.

\subsection{Flat solutions of the leading large $d$ equations}

It is obvious that constant $G_{ab}$ is a solution to \eqref{ldeom}
provided 
\begin{equation}\label{fsssol}
 e^{\phi} \chi^2=4
\end{equation}
(here $\chi^2= G^{ab} \chi_a \chi_b$)
\eqref{fsssol} is equivalent to (see \eqref{cc})
\begin{equation} \label{phid}
e^{\phi} (g^{\mu\nu} \partial_\mu \phi  \partial_\nu \phi)=4
\end{equation} 
Its easy to check that \eqref{phid} is satisfied at every point in (global) 
flat space. In the coordinate system of \eqref{fss} this follows 
upon noting that  $\phi=2 \ln S$ and $e^\phi=S^2$.

\subsection{The black brane}

In this subsection we present a less trivial class of exact solutions to the 
leading order large $d$ equations of motion \eqref{ldeom}. The solutions 
we present in this subsection maintain translational invariance in 
$p+2$ out of the $p+3$ dimensions in which we work; we refer to them 
as black brane solutions (in analogy 
with the black branes of asymptotically $AdS$ Einstein gravity with a 
negative cosmological constant). 

Let $G_F$ denote any metric whose components, $(G_F)_{ab}$ are all constants. 
Consider the metric
\begin{equation}\label{bbf}
G_F+ e^{-R} O^2
\end{equation}
where $R$ is a coordinate ($dR$ is the direction in which translational 
invariance is broken) and  $O$ is a constant one form (i.e. each component of 
$O_\mu$ is a constant). It turns out that \eqref{bbf} is an exact 
solution to the equation of motion \eqref{ldeom} provided that 
\begin{equation}\label{conditions}
e^{\phi} \chi. \chi=4 , ~~~(2 dR -\chi). dR=0, ~~~(2dR-\chi).O=0, ~~~
O.O=0
\end{equation}
(all dot products in equation \eqref{conditions} are evaluated 
using the metric $G_F$). \footnote{Note that the metric \eqref{bbf}
does not provide a completely unambiguous definition of the coordinate $R$
and the one form $O$. The redefinition $R \rightarrow R+ m$ and 
$O \rightarrow e^{\frac{m}{2}}  O$ leaves \eqref{bbf} unchanged. It follows that 
the one form $O$ and the coordinate $R$ are defined only upto this ambiguity.}
In Appendix \ref{bb} we have demonstrated that it is possible to choose 
coordinates in which the metric \eqref{bbf} takes the form 
\begin{equation}\label{pea} \begin{split}
&e^{-\phi}=\frac{1}{x_0^2}, ~~~\chi=2 d R + \frac{2}{x_0} dX \\
&ds^2= 2 dR dV - a \left(1-e^{-R} \right) dV^2 + dY^i dY^i + \frac{dX^2}{1-a x_0^2}
\\
\end{split}
\end{equation}
 At fixed $x_0$ it follows that 
\eqref{bbf} and \eqref{conditions} define a one parameter set of metrics 
upto coordinate transformations. In Appendix \ref{bb} we demonstrate that 
the one parameter, $a$ in \eqref{pea}, is a consequence of the scaling 
symmetry  \footnote{Namely that uniform
scaling of the metric by a constant maps solutions to solutions.}
of the vacuum Einstein equations without a cosmological constant.

The black brane solutions presented in this subsection are not really 
new. They can be obtained from the well known Schwarzschild solution
\eqref{schbh} as follows. We choose any point on the horizon of the black 
hole, and then zoom into a patch of size $\frac{1}{d}$ of that point by 
performing the coordinate changes \eqref{cc}. The resultant metric turns out 
to be the black brane; the parameter $a$ of the black brane is a function of 
$r_0$ of the original black hole. In other words the black brane solutions 
of this section are simply large $d$ limits of appropriately `boosted' 
versions of the `near horizon' solutions of Schwarzschild black holes 
(of various sizes).

\subsection{Sewing black branes into a membrane}

In this subsection we will present the construction of a class of metrics 
with several interesting properties. We start with flat space and perturb
the metric in a manner controlled by a scalar function $B$ and a null vector 
field $O$.

Let $B$ be an $SO(d+1)$ invariant scalar function that lives on the 
flat space \eqref{fss}.  $SO(d+1)$ invariance ensures that $B$ is 
a function only of the coordinates $(w^a, S)$ (see around \eqref{fss} 
for definitions). We choose the function $B$ so that its zeroes 
form a codimension one closed surface -for instance of the topology of a sphere 
times time -  in the flat spacetime \eqref{fss}. We refer to this surface as the membrane. We require that $B$ have only simple zeroes. The zeroes of 
$B$ divide 
flat space up into two disjoint regions; we refer to the part of spacetime 
that includes infinity as the outside, and the other region as the inside. 
We require that $B$ is negative on the inside and positive on the outside, 
and also that $dB. dS>0$. 

In terms of $B$ we define an auxiliary scalar 
function $\psi$ by the equation
\begin{equation}\label{psidef}
\psi=1+ \frac{dB.d\phi}{2 dB.dB}B
\end{equation}
It follows from the conditions on $B$ that $\psi<1$ inside, that 
$\psi=1$ on the membrane and that $\psi>1$ on the outside. 

In addition to $B$, let us consider a null $SO(d+1)$ invariant 
one form field $O$ which lives in 
flat space. As $O$ is null, $O.O=0$ everywhere. We also demand that 
\begin{equation}\label{psi_normalisation}
 {O}.(\frac{d\phi}{2}-d\psi)|_{B=0}=0. 
\end{equation}
where \eqref{psi_normalisation} is required to hold only on the membrane. 
The functions $B$ and $O$, written as functions of $(w^a, S)$ and $(dw^a, dS)$,
are assumed to be independent of $d$.

Now consider the spacetime 
\begin{equation}\label{fssa} 
ds^2=ds_{flat}^2 + \frac{O_\mu O_\nu dx^\mu dx^\nu}{\psi^{d+p}}
\end{equation}
Here $ds_{flat}^2$ is the metric of flat space (for instance 
$ds_{flat}^2$ is given by \eqref{fss} in a useful coordinate system).

The spacetime \eqref{fssa} has several interesting properties. First, it 
approaches the metric of flat space exponentially rapidly in the 
outside region. The deviation of \eqref{fssa} from flat space scales like 
$e^{-(\psi-1)d}$. It follows, in particular, that \eqref{fssa} solves 
Einstein's equations at large $d$ at any point outside the membrane 
such that $\lim_{d \to \infty }(\psi-1)d = \infty$. 

The only points outside the membrane at which it is 
not obvious that \eqref{fssa} solves Einstein's equations is points for which 
$\psi-1$ is scaled to be of order $\frac{1}{d}$ as $d$ is taken to infinity. 
Recall that $\psi=1$ defines the membrane surface. As $\psi$ is a smooth function
in the neighbourhood of the membrane, all such points lie within a distance $\frac{1}{d}$ 
of the membrane. We refer to this neighbourhood of the membrane as the 
`membrane region'. We will now argue that \eqref{fssa} solves the leading 
order large $d$ Einstein equations everywhere within the membrane region.

In the membrane region $\psi^{p+d} \sim e^{(p+d)(\psi-1)}$. As $\psi$ is 
a smooth regular $d$ independent function in the neighbourhood of the membrane, 
it follows that $e^{(p+d)(\psi-1)}$, and so the metric \eqref{fssa}, 
varies over a length scale of order $\frac{1}{d}$. 
In contrast $\phi=2 \ln S$ varies over the length scale unity.
 
Let us now study the metric in the membrane region in more detail, using the 
strategy outlined around \eqref{cc}. For this purpose it is useful to use 
$\psi-1$ as one of our original coordinates, and choose $R=d(\psi-1)$ as the corresponding 
rescaled coordinate for the blown up analysis of \eqref{cc}. Choosing $x_0^\mu$ 
in \eqref{cc} to be any point on the membrane, it is 
easily seen that \eqref{fssa} reduces precisely to the black brane metric 
\eqref{bbf} at leading order in the large $d$ limit. As the flat space metric and $O$ were smooth functions in the 
original coordinates, they are constants after blowing up (to leading order in $\frac{1}{d}$)
exactly as for \eqref{bbf}. Moreover all the conditions \eqref{conditions} are also easily 
seen to be obeyed. The first of 
\eqref{conditions} follows from the fact that $\phi=2 \ln S$. The second 
equation follows upon using \eqref{psidef} and the fact that $B$ vanishes
on the membrane. The third of \eqref{conditions} follows directly from 
\eqref{psi_normalisation} and the fourth of \eqref{conditions} follows 
form the fact that $O$ is a null field. 

Finally, in  Appendix \ref{eheh} we demonstrate that the event horizon 
of the metric \eqref{fssa} lies within the membrane region. For this 
reason we do not care whether \eqref{fssa} obeys Einstein's equations 
inside the membrane (i.e. for $1-\psi \gg \frac{1}{d}$) as this region of 
spacetime is causally disconnected from the outside.

In summary the metric 
\eqref{fssa} obeys the leading large $d$ Einstein equations everywhere
outside its event horizon. This fact allows us to use \eqref{fssa}
as the starting point for the construction of solutions to 
Einstein's equations in a systematic  expansion in $\frac{1}{d}$. 

\subsection{Equivalent starting points for perturbation theory and a 
`gauge choice'}

Recall that the metric \eqref{fssa} obeys Einstein's equations only at leading
order in large $d$. If two metrics of the form \eqref{fssa} coincide 
at leading order in large $d$ (but differ at subleading order) they should
be regarded as equivalent starting points for perturbation theory. 
With this definition of equivalence classes we  
show in Appendix \ref{inequiv}  that  inequivalent classes of metrics 
\eqref{fssa} are labelled
by the shape of the membrane and the value of the one form field $O$ {\it on} 
the membrane. In other words functions $B$ and one forms $O$ related by 
\begin{equation}\label{sred}
O \rightarrow O+ \delta O, ~~~B \rightarrow \alpha B
\end{equation}
are equivalent starting points for perturbation theory. \footnote{
We assume $\delta O$ vanishes on the membrane while $\alpha$ is nonzero but
finite on the membrane.}

In order to proceed with our perturbative construction of metrics 
in the large $d$ limit, we will find it useful to choose one representative 
in each equivalence class; we do this by imposing the following  
`gauge' conditions on $B$ and $O$
\begin{equation} \label{gcop} \begin{split}
&(\partial B. \nabla) \nabla_\mu B=0\\
& \partial B. \nabla  O=0\\
\end{split}
\end{equation}
The first condition in equation \eqref{gcop} 
asserts that $\nabla_\mu B$ obeys the geodesic
equation with affine parametrization. The second condition in \eqref{gcop} 
asserts that $O$ is simply parallel transported along these geodesics.

It is useful to note a simple consequence of \eqref{gcop}
\begin{equation}\label{constnorm}
\nabla_\alpha (\partial_\mu B \partial^\mu B)
= 2 {\partial^\mu B} \nabla_\mu  (\nabla_\alpha B)= 0 .
\end{equation}
In other words the norm of the one form $dB$ is a constant - not just along the 
geodesics, but everywhere in space. An overall scaling of $B$ can then be 
used to set the norm of this one form to unity. With this choice of scaling, 
$dB$ is, in particular, the unit normal one form to the membrane. We employ 
this choice in what follows. \footnote{With these conventions there is 
an elegant geometrical construction of the function $B$ in the neighbourhood
of the membrane. Let us foliate spacetime with geodesics that pierce the 
membrane in a normal direction. The value of $B$ at any point $P$ is simply 
equal to the signed length along the appropriate geodesic from the membrane 
to $P$.} In the rest of this paper we use the notation 
\begin{equation} \label{ndef}
n_\mu(x)= \partial_\mu B
\end{equation}
which highlights the fact that $\partial_\mu B$ is the unit normal vector on the 
membrane. In this new notation \eqref{gcop} become 
\begin{equation}\label{gcopn}
n. \nabla n = n.\nabla O =0
\end{equation}

Finally let us turn to a more explicit parametrization of the one form 
$O$. Let 
\begin{equation}\label{param}
O=e^h (1, -u_a)
\end{equation}
(where the first component in $(1, -u_a)$ is the coefficient of $dS$, 
and the remaining components refer to the components of $dw^a$. It follows 
from the fact that $O$ is null that 
\begin{equation}\label{unitnorm}
u^2=-1
\end{equation}
Moreover, the second of \eqref{gcop} implies that 
\begin{equation}\label{ngc}
n. \nabla u_{a} = n. \nabla h =0
\end{equation}
Finally recall that $O. (d \psi- \frac{dS}{S})=0$ on the membrane. Using 
$\partial_\mu\psi= \frac{\partial_S B \partial_\mu B}{S}$ on the membrane, 
 this condition reduces on the membrane to 
\begin{equation}\label{vcons}
u.\partial B= \partial_S B - \frac{1}{\partial_S B}
\end{equation}
Now it is easily verified that the 
$\nabla^\mu B\nabla_\mu$ derivative of the equation \eqref{vcons} vanishes 
(this follows from the second of \eqref{gcop}) implying that 
\eqref{vcons} holds everywhere, and not just on the membrane. 

\section{Correction at first order in $\frac{1}{d}$}

In this section we will improve the metric \eqref{fssa} to ensure 
that it solves Einstein's equations at first subleading order in 
$\frac{1}{d}$. Our strategy to do this is straightforward. 
We will expand the metric \eqref{fssa} about  $x_0^\mu$ in the patch coordinates
\eqref{cc}, but will keep both the leading and first subleading terms in 
the $\frac{1}{d}$ expansion. In an appropriate choice of patch coordinates 
the metric reduces to the black brane solution \eqref{pea} corrected 
by terms of order $\frac{1}{d}$ that are linear in 
$\delta x^\mu=x^\mu-x_0^\mu$. This metric fails to solve Einstein's equations
at first subleading order in $\frac{1}{d}$. In order to fix this defect 
we allow the metric \eqref{fssa} to be corrected at first order in
$\frac{1}{d}$ and determine the form of the correction that ensures that 
Einstein's equations are obeyed at first order in $\frac{1}{d}$. As we 
will see below, the correction is nonsingular if and only if 
the membrane shape and $u_\mu$ field obey 
the equations of motion \eqref{mainequations}.

\subsection{Interpretation of Taylor coefficients and constraints} \label{te}

The functions $B(x)$, $h(x)$ and $u_\mu(x)$ can be Taylor expanded about any 
point $x_0^\mu$ as follows
\begin{equation}\label{texp} \begin{split}
B(x)&= B(x_0)+ \nabla_\mu B (x_0) \delta x^\mu + 
\nabla_\mu \nabla_\nu B(x_0) \frac{ \delta x^\mu \delta x^\nu}{2} + \ldots\\
h(x)&=h(x_0)+ \nabla_\mu h(x_0) \delta x^\mu + \ldots \\
u_\mu(x)&=u_\mu(x_0) +\nabla_\nu u_\mu (x_0) \delta x^\nu + \ldots \\
\end{split}
\end{equation}
where $\delta x^\mu= x^\mu - x_0^\mu$. Let us define the functions 
\begin{equation}\label{fns}
\nabla_\mu B (x)=n_\mu(x), ~~~\nabla_\mu \nabla_\nu B(x)=K_{\mu\nu}(x), 
~~~\nabla_\mu h(x)=h_\mu(x), ~~~\nabla_\nu u_\mu (x)=u_{\mu\nu}(x)
\end{equation}
It follows that the Taylor expansion \eqref{texp} may be rewritten as 
\begin{equation}\label{texp} \begin{split}
B(x)&= B(x_0)+ n_\mu (x_0) \delta x^\mu + K_{\mu\nu} 
\frac{ \delta x^\mu \delta x^\nu}{2} + \ldots\\
h(x)&=h(x_0)+ h_\mu(x_0) \delta x^\mu + \ldots \\
u_\mu(x)&=u_\mu(x_0) + u_{\mu\nu} (x_0) \delta x^\nu + \ldots \\
n_\mu(x)&=n_\mu(x_0) + K_{\mu\nu}(x_0) \delta x^\nu + \ldots \\
\end{split}
\end{equation}
Recall that the norm of $n_\mu(x)$ is a constant (see \eqref{constnorm}) 
and that we have chosen to normalize $B$ to set this constant to unity. 
It follows from the first of \eqref{gcop} that 
\begin{equation} \label{orthog}
n^\mu(x) K_{\mu\nu}(x)=0
\end{equation}
Using \eqref{texp} and \eqref{orthog}, it follows that the extrinsic 
curvature at any point $x_0^\mu$ on the membrane surface is given 
by simply $K_{\mu\nu}(x_0)$.

The derivatives of the velocity field are not all independent; differentiating
\eqref{vcons} we find   
\begin{equation}\label{fdvc}
n^\mu u_{\mu\nu}= 
\left( -K_{\nu\mu}u^\mu +K_{\nu s} + \frac{1}{n_s^2} K_{\nu s} \right)
\end{equation}
and differentiating \eqref{unitnorm} we have
\begin{equation}\label{uorthog}
u^\mu u_{\mu \nu}=0
\end{equation}

\subsection{Data}

At any point on the membrane, there is a $p$ dimensional subspace orthogonal 
to the three dimensional subspace spanned by $n_\mu, u_\mu, dS$. In this 
brief subsection we list a basis of independent Taylor expansion coefficients, 
labelled by their transformation properties under the local tangent 
space $SO(p)$ group. Recall that the  Taylor expansion coefficients 
$K_{\mu\nu}$, $u_{\mu\nu}$ and $h_\mu$ are not independent as they are 
constrained by \eqref{orthog}, \eqref{uorthog},\eqref{ngc} and \eqref{fdvc}; 
these equations can be used to determine all the `other' expansion 
coefficients in terms of the independent coefficients listed below
\begin{itemize}
\item Scalars:
\begin{equation*}
\begin{split}
&S_1=K_{ss},~S_2=u^\mu K_{s\mu},~S_3=u^\mu u^\nu K_{\mu\nu},~S_4={\cal P}^{\mu\nu}K_{\mu\nu}\\
&S_5=h_s,~S_6=u^\mu h_\mu,~S_7={\cal P}^{\mu\nu}u_{\mu\nu}
\end{split}
\end{equation*}
\item Vectors:
$$V^\mu_1={\cal P}^{\mu\alpha}u_{\alpha s},~V^\mu_2={\cal P}^{\mu\alpha}u^\nu u_{{\alpha\nu}},~V^\mu_3={\cal P}^{\mu\alpha}K_{s\alpha},~V^\mu_4={\cal P}^{\mu\alpha}u^\nu K_{\nu\alpha},~V^\mu_5={\cal P}^{\mu\alpha}h_{\alpha}$$
\item Symmetric Tensor:
$$T^{\mu\nu}_1={\cal P}^{\mu\alpha} P^{\mu\beta}\left[ K_{\alpha\beta} - \frac{\eta_{\alpha\beta}}{p}{\cal P}^{\theta\phi}K_{\theta\phi}\right],~T^{\mu\nu}_2={\cal P}^{\mu\alpha} P^{\mu\beta}\left[\frac{ u_{\alpha\beta} +u_{\beta\alpha} }{2}- \frac{\eta_{\alpha\beta}}{p}\left({\cal P}^{\theta\phi}u_{\theta\phi}\right)\right]$$
\item Anti-symmetric tensor:
$$A^{\mu\nu}={\cal P}^{\mu\alpha}P^{\nu\beta}\left[\frac{ u_{\alpha\beta}{ -}u_{\beta\alpha} }{2}\right]$$
\end{itemize}

As we have explained above, our `gauge' condition \eqref{gcop} ensures that normal derivatives 
of $h$ and the velocity field vanish (see \eqref{ngc}). This ensures that the derivatives
of these quantities are equal to derivatives projected to the world volume of the membrane, i.e. 
\begin{equation}\label{equivproj}
u.\nabla  = u.\nabla - (n.u) n. \nabla, ~~~\partial_S= \partial_S-n_S (n. \nabla)
\end{equation}
As the $h$ and velocity fields are physical only on the surface of the membrane, 
all derivatives in the data above (and in the final first order metric for our black hole below) 
should actually be thought of as projected derivatives. With this replacement our 
final results for the metric and equation of motion apply without reference to the arbitrary 
gauge choice \eqref{gcop}.

\subsection{Structure of the first correction to the metric}

As mentioned above, we will need to correct the metric \eqref{fssa} in order 
to ensure that Einstein's equations are obeyed at first subleading order 
in $\frac{1}{d}$. We adopt $S$, the size of the $d$ sphere as one of our 
coordinates. This choice leaves us with $p+2$ remaining coordinate freedoms. 
We fix this freedom in the following manner. Let $d\psi$, $dS-S d\psi$, 
$O$ and ${\cal P}^\mu_\nu dx^\nu$ form a basis of the $p+3$ 
dimensional space of one forms at any point. 
The correction metric can be regarded, 
point wise, as a symmetric quadratic form of these one forms. Our choice of 
gauge is that the only terms involving $d \psi$ in the correction metric 
are of the form 
$$d\psi \otimes (d\psi -\frac{dS}{S});$$
\footnote{We use the symbol $\otimes$ to denote the symmetrized product of 
two one forms. If $A= A_\mu dx^\mu$ and $B=B_\mu dx^\mu$ then $A \otimes B = A_\mu B_\nu dx^\mu dx^\nu$. In our convention $A^2=A \otimes A$.}
The correction metric will not contain terms proportional to 
$d\psi \otimes d\psi$, $d\psi \otimes O$ and 
$d\psi \otimes  {\cal P}^\mu_\nu dx^\nu$. 

With this choice of gauge the metric \eqref{fssa}, corrected to first 
order in $\frac{1}{d}$, must take the form 
\begin{equation}\label{prathomic}
\begin{split}
ds^2 &= \eta_{\mu \nu}dx^\mu dx^\nu + \psi^{-(d+p)} O^2\\
&+\frac{\psi^{-d} }{d}\bigg[K_1 (x^\alpha) O^2 + 2K_2(x^\alpha)(dS -Sd\psi)\otimes O\\
&~~~~~~~+ K_3(x^\alpha)(dS -Sd\psi)^2 +2 K_4(x^\alpha)d\psi\otimes (dS -Sd\psi)\\
&~~~~~~~+2Q^\beta_1(x^\alpha) O_\mu{\cal P}_\nu^\beta dx^\mu dx^\nu+2Q^\beta_2(x^\alpha) {\cal P}_\nu^\beta dx^\nu(dS -Sd\psi)\\
&~~~~~~~+{\cal T}_{\alpha\beta}(x^\alpha){\cal P}_\mu^\alpha {\cal P}_\nu^\beta dx^\mu dx^\nu \bigg]
+{\cal O}\left(\frac{1}{d}\right)^2.
\end{split}
\end{equation}

The ${\cal O}(\frac{1}{d})$ correction piece in \eqref{prathomic} has been 
added to ensure that \eqref{prathomic} solves Einstein's equations at 
first subleading order in $\frac{1}{d}$. This correction is needed because 
the first line of \eqref{prathomic} does not, by itself, have this property. 
Let us recall why this is the case. Upon expanding the first line of 
\eqref{prathomic} in a patch to first subleading order in $\frac{1}{d}$,
we find the black brane background corrected by ${\cal O}(1/d)$ fluctuations. 
Now the fluctuations - which are all proportional to the data of the 
previous subsection plus a constant piece \footnote{The constant arises from 
the fact that $\frac{1}{\psi^{p+d}}$ reduces to $e^{-R}$ only at leading order 
at large $d$; there are corrections to this formula at first subleading order in
$\frac{1}{d}$.}- fail to solve the leading order Einstein equations, 
giving rise to sources proportional to first order data. Moreover 
the black brane metric itself is a solution only to the leading order 
Einstein equations, and so yields a constant source at order $\frac{1}{d}$ 
when plugged into the exact Einstein equations. As the correction part of 
the metric in \eqref{prathomic} is chosen to cancel these sources, it follows 
that the unknown functions in \eqref{prathomic} must take the form
\begin{equation}\label{bistar}
\begin{split}
K_a(x^\alpha) &= \sum_{A=0}^7 K_a^A(R) S_A,~~a=\{1,2,3,4\},~~S_0 =1\\
Q^\mu_a(x^\alpha) &= \sum_{A=1}^5Q_a^A(R) V^\mu_A,~~a=\{1,2\}\\
{\cal T}^{\mu\nu}(x^\alpha) &=\sum_{A=1}^2{\cal T}_A(R) T^{\mu\nu}_A\\
\text{where} ~~~R&= d(\psi-1)
\end{split}
\end{equation}
\footnote{The appearance of $S_0=1$ is a consequence of the fact that we have constant 
sources (not proportional to the first order data of the previous subsection) 
as explained above.} The unknown functions $K_a^A(R)$, $Q_a^A(R)$, ${\cal T}_A(R)$ 
are now functions only of $R$ and can be regarded as constants in the 
other black brane directions within a patch (their variation in these 
directions affects the metric only at ${\cal O}(1/d^2)$). 

\subsection{Results}

In order to determine the unknown functions in \eqref{bistar} we 
 now expand the metric \eqref{prathomic} in patch centred about a particular 
point $x_0^\mu$ and obtain the patch metric expanded to first order in 
$\frac{1}{d}$
(including terms that come from the Taylor expansion of the first line). 
We then move to a convenient set of patch coordinates and plug the resultant 
metric into Einstein's equations. As we explain in detail in Appendix 
\ref{method}, we find differential equations for all unknown functions 
in \eqref{bistar} 
which we proceed to solve. The solution to these equations turns out to 
include a singularity within the membrane region (see Appendix \ref{method})
unless the equations \eqref{mainequations} are obeyed. When those equations
are obeyed the unknown functions in \eqref{bistar} are all quite simple 
and we find \footnote{In this paper we have performed all our computations 
only for the special case $p=2$. The formulae presented in equations 
\eqref{maap1a}, \eqref{maap1b} and \eqref{maap1c} are accurate only for 
$p=2$. We do not, however, need to perform a new computation in order to 
obtain the analogues of these three functions at arbitrary $p$. This is 
because a solution at $p=2$ may be viewed special case of a 
solution of the membrane equations at all larger values of $p$ - simply by 
moving some of the $z_M$ coordinates into the set of $w_a$ coordinates. 
This operation changes the meaning of $S$, and so induces a non-trivial 
redefinition of the velocity one form $u_\mu$ (recall $u$ is defined so 
that $O \propto (dS -u)$). This requirement of `redistribution invariance'
 together with the explicit $p=2$ expressions presented below in 
\eqref{maap1a}, \eqref{maap1b} and \eqref{maap1c}
is sufficient to completely fix the three functions $K_1$, $K_2$ and $K_3$ 
at all $p$. We postpone the presentation of the explicit expression 
for the first order correction to the membrane metric at arbitrary values of 
$p$ to an upcoming paper. Similar logic may be used 
to constrain the $p$ dependence of the membrane equations of motion 
\eqref{mainequations}. In this case we find that the requirement 
of redistribution invariance infact forces the membrane equations to take 
the form \eqref{mainequations} at all $p$. This fact will also be explained 
more fully the upcoming paper mentioned above.} 
that the final metric (\eqref{prathomic} with explicit solutions for the functions in \eqref{bistar}) is given by 

\begin{equation}\label{maap1}
\begin{split}
ds^2 &= \eta_{\mu \nu}dx^\mu dx^\nu + \psi^{-(d+p)} (O_\mu dx^\mu)^2\\
&+\frac{ \psi^{-d} }{d}(O_\mu dx^\mu)\bigg[K_1 (x^\alpha) (O_\nu dx^\nu)+ 2K_2(x^\alpha)(dS -Sd\psi)
-K_V(x^\alpha) ({\cal P}_\nu^\beta u_{\beta s })dx^\nu\bigg]\\
&+{\cal O}\left(\frac{1}{d}\right)^2\\
\end{split}
\end{equation}
where
\begin{equation}\label{maap1a}
\begin{split}
K_1 (x^\alpha)&=-\left[\frac{-c^6+11 c^4-23 c^2+11}{\left(c^2-2\right)^2 \left(c^2-1\right)}\right] R- \left[\frac{\left(c^2-1\right) \left(3 c^2-8\right)}{2 \left(c^2-2\right)^2}\right]R^2\\
&+\left[\frac{c^2 R^2+2 \left(-4 c^4+7 c^2+1\right) R}{2 c^2 \left(2-c^2\right)^2}\right](u^\alpha u^\beta K_{\alpha\beta})(S n_S)\\
&-\left[\frac{c^2 R^2+\left(-4 c^4+6 c^2+2\right) R}{c^2\left(1-c^2\right)
   \left(2-c^2\right)}\right]( u^\beta K_{S\beta} )(S n_S)\\
&   +\frac{R}{1-c^2}\left[2S\partial_S h- ({\cal P}^{\mu\nu}K_{\mu\nu})(S n_S)\right]\\
\end{split}
\end{equation}
\begin{equation}\label{maap1b}
\begin{split}
K_2(x^\alpha) &=e^h\left[\frac{(3 c^4-6 c^2+1)+\left(c^6-3 c^4+2 c^2\right) R}{c^2 \left(c^2-2\right)^2
   \left(c^2-1\right)}\right]\\
   &+e^h\left[\frac{(3 c^4-6 c^2+1)+\left(c^2-2\right) c^2 R}{c^4 \left(2-c^2\right)^2}\right](u^\alpha u^\beta K_{\alpha\beta})(S n_S)\\
  & -e^h\left[\frac{-\left(2-c^2\right) R+3 c^2-5}{c^2 \left(1-c^2\right) \left(2-c^2\right)}\right]( u^\beta K_{S\beta} )(S n_S)\\
 & -\frac{S e^h}{c^2(1-c^2)}\bigg[2(2-c^2)(\partial_S h) - 2(1-c^2)(u^\theta\partial_\theta h)\\
 &~~~~~~~~~~~~~~~~~~~-({\cal P}^{\mu\nu}K_{\mu\nu})( n_ S) + (1-c^2)(P^{\mu\nu}u_{\mu\nu})\bigg]\\
 \end{split}
 \end{equation}
 \begin{equation}\label{maap1c}
 \begin{split}
 K_V(x^\alpha) &=  \frac{2Se^h}{(1-c^2)}(1+R).
\end{split}
\end{equation}
Here 
\begin{equation}\label{maap3}
\begin{split}
{\cal P}_{\mu\nu} &= \text{The projector perpendicular to $u_\mu$, $n_\mu$ and $dS$}\\
c^2 &= 1-n_S^2\\
R&= d(\psi -1)
\end{split}
\end{equation}
All raising and lowering of the indices have been done using the flat metric $\eta_{\mu\nu}$. 

\eqref{maap1} is our final result for the metric of membrane spacetimes 
corrected to first order in $\frac{1}{d}$. We recall that \eqref{maap1}
solves Einstein's equations to the relevant order if and only if 
\eqref{mainequations} are obeyed.

\section{Stationary black hole solutions}
In this section we provide some examples of exact solutions to Einstein's 
equations that reduce to membrane spacetimes at large $d$. We explicitly 
verify that these solutions obey the equations \eqref{mainequations}.
\subsection{The Schwarzschild black hole}
The metric for the Schwarzschild black hole of unit radius is given by
\begin{equation}
ds^2=-dt^2+dS^2+dx_idx^i+\left(dt+\frac{x_i dx^i+SdS}{\sqrt{r^2+S^2}}\right)^2\frac{1}
{\left(S^2+r^2 \right)^{\frac{p+d}{2}}}
\end{equation}
Hence the zero norm one form $O$ is given by
\begin{eqnarray}
O&=&\frac{S}{\sqrt{S^2+r^2}}(dS-u_{\mu}dx^{\mu})\nonumber\\
u_{\mu}dx^{\mu}&=& -\frac{\sqrt{S^2+r^2}}{S}dt-\frac{r}{S}dr
\end{eqnarray}
where $r^2= x_i x^i$.  The membrane surface is defined by
\begin{equation}
S^2+x_ix^i=1
\end{equation}
the normal to the surface which has unit magnitude on the surface is given by
\begin{equation}
n=\frac{SdS+rdr}{\sqrt{S^2+r^2}}
\end{equation}
We see that the conditions $u.u=-1$ and $u.n=n^s-\frac{n^2}{n^s}$ are trivially satisfied. 

Let us now focus on the membrane surface $S^2+r^2=1$. 
The non-zero components of extrinsic curvature of this surface are obtained as
\begin{eqnarray}
K_{SS}&=& 1-S^2\nonumber\\
K_{Si}&=&-S x_i=K_{iS}\nonumber\\
K_{ij}&=& \delta_{ij}-x_ix_j
\end{eqnarray}

We will now check that the Schwarzschild black hole obeys the equations 
\eqref{mainequations}. The value of the constant $c^2$ in these equations 
 turns out to be $r^2$ when evaluated on this solution. 
 The various terms appearing in the scalar equations evaluate to
\begin{equation}
(2-c^2)^2K_{SS}=(1+S^2)^2r^2,\quad (1-c^2)^2K_{uu}=S^4r^2,\quad 2(1-c^2)(2-c^2)K_{Su}=2(1-S^2)S^2r^2
\end{equation}
Plugging all this in the scalar equations of \eqref{mainequations} we see that it is satisfied. The two terms in the vector equations  of \eqref{mainequations} individually evaluate to zero on the membrane and hence it is also satisfied. 

In conclusion, the Schwarzschild black hole spacetime takes the membrane form 
at large $d$; the shape of the corresponding membrane and its velocity field 
solve the membrane equations of motion.

\subsection{The Myers-Perry black hole}

Consider the Myers-Perry black hole \cite{Myers:1986un,Myers:2011yc}
with independent rotations in $q$ 
distinct two planes.
In order to describe this solution within our formalism we need to choose 
$p+1 \geq 2q$. We find it convenient to saturate this inequality so that 
$p+1=2q$ or $q= \frac{p+1}{2}$.  

In Appendix \ref{mp} we demonstrate that the Myers-Perry black hole with 
rotations in $\frac{p+1}{2}$ planes is described by the metric 

\begin{equation}\label{bhmet}
ds^2=-dt^2+\sum_{i=1}^{\frac{p+1}{2}}(dx_i^2+dy_i^2)+dS^2+\frac{O^2}{\rho^{d+p}}
\end{equation}
where $S^2= z_M z^M$ as around \eqref{fss} and 
\begin{eqnarray}
O&=& A\left(dt+\left(1-\sum_{i=1}^{\frac{p+1}{2}}\frac{\mu_i^2a_i^2}{\rho^2+a_i^2}\right)d\rho-\sum_{i=1}^{\frac{p+1}{2}}\frac{a_i}{\rho^2+a_i^2}(x_idy_i-y_idx_i)\right)\quad \mbox \nonumber\\
A&=&\left( m \frac{1}{\prod_{i=1}^\frac{p+1}{2} (1+\frac{a_i^2}{\rho^2}) }
\frac{1}{ \left(1-\sum_{i=1}^{\frac{p+1}{2}}\frac{a_i^2\mu_i^2}{\rho^2+a_i^2}\right)}\right)^{\frac{1}{2}}.
\end{eqnarray}
and 
\begin{equation} \begin{split} \label{defmr}
&\sum_{i=1}^{\frac{p+1}{2}}\frac{x_i^2+y_i^2}{\rho^2+a_i^2} + \frac{S^2}{\rho^2} = 1.\\
&\mu_i^2 = \frac{x_i^2+y_i^2}{\rho^2+a_i^2}.
\end{split}
\end{equation}
so that 
\begin{equation}\label{drhodef} 
\rho d\rho=\frac{1}{\sum_{j=1}^{\frac{p+1}{2}}\frac{x_j^2+y_j^2}{(\rho^2+a_j^2)^2}+\frac{S^2}{\rho^4}}\left(\sum_{i=1}^{\frac{p+1}{2}}\frac{x_idx_i+y_idy_i}{\rho^2+a_i^2}+\frac{S}{\rho^2}dS\right)\quad 
\end{equation} 
The metric \eqref{bhmet} takes the form \eqref{fssa} with $\psi=\rho$ 
(recall $\rho$ is defined implicitly by \eqref{defmr}). It is not difficult 
to verify that $O_\mu O^\mu=0$ on the membrane $\rho=1$. It is not difficult 
to show that 
$$ O \propto dS-u_{\mu}dx^{\mu}$$
where
\begin{eqnarray} \label{urot}
u&=&-\frac{1}{\xi}\left(dt+\sum_{i=1}^{\frac{p+1}{2}}\left(\frac{x_i+a_i y_i}{1+a_i^2}dx_i+\frac{y_i-a_ix_i}{1+a_i^2}dy_i\right)\right)\nonumber\\
\xi&=& 1-\sum_{i=1}^{\frac{p+1}{2}}\left(\frac{x_i^2+y_i^2}{1+a_i^2}\right)
\end{eqnarray}
It is easily verified that $u_\mu u^\mu=-1$. 
The unit one form normal to the membrane is proportional to $d \rho$; upon 
normalizing we find  
\begin{equation} \label{nrot}
n=\frac{1}{\sum_{i=1}^{\frac{p+1}{2}}\frac{x_i^2+y_i^2}{(1+a_i^2)^2}+S^2}\left(\sum_{i=1}^{\frac{p+1}{2}}\frac{x_idx^i+y_idy^i}{1+a_i^2}+S dS\right)
\end{equation}
We have explicitly checked that 
$$u.n=n_S-\frac{n.n}{n_S}$$
so that \eqref{vcons} is obeyed. It follows that the metric \eqref{bhmet} agrees
with our form for the membrane metric at leading order in large $d$. 

We have also verified that the expressions \eqref{urot} and \eqref{nrot} 
obey the equations of motion \eqref{mainequations}. In order to verify the 
scalar equation (first of \eqref{mainequations}) we explicitly
 computed the extrinsic curvature tensor. Using our explicit (and lengthy) 
expressions, we find that the scalar equation (first of \eqref{mainequations}) 
is indeed satisfied. 

We have also verified that the vector equation (the second of 
\eqref{mainequations} is obeyed. Let us specialize to the case $q=1$ or 
$p=1$. On the membrane surface we find 
\begin{eqnarray}
{\cal P}_a^b \left( (n_S^2) \left( u. \nabla   - (u.n)   
n. \nabla\right) u_b\right)&=&\left(\begin{matrix}
0\\\frac{a^2 \left(a^2+1\right) \left(x^2+y^2\right) \left(2 a^4-2 a^2
	\left(x^2+y^2-2\right)-x^2-y^2+2\right)}{\left(a^4-a^2 \left(x^2+y^2-2\right)+1\right)^3}\\\frac{a \left(a^2+1\right)^2 y \left(2 a^4-2 a^2 \left(x^2+y^2-2\right)-x^2-y^2+2\right)}{\left(a^4-a^2 \left(x^2+y^2-2\right)+1\right)^3}\\-\frac{a \left(a^2+1\right)^2 x \left(2 a^4-2 a^2 \left(x^2+y^2-2\right)-x^2-y^2+2\right)}{\left(a^4-a^2 \left(x^2+y^2-2\right)+1\right)^3}\end{matrix}\right)\nonumber\\
&=&{\cal P}_a^b \left(( 1+n_S^2 )
\left( \partial_s - n_S n. \nabla \right) u_b \right)
\end{eqnarray} 
(where we have listed the $S$, $t$, $x$ and $y$ components of the vector 
in that order). It follows that the second of \eqref{mainequations} is 
satisfied. Note that each individual term in this equations is non-zero; 
however the two terms cancel between each other. Although we do not explicitly 
report our results here, we have also verified that the vector equation is satisfied
for the case $q=2$ or $p=3$. 

\section{Linearization of membrane equations about the Schwarzschild solution}

In the previous section we verified that rotating black holes at large $d$ 
are solutions to our membrane equations. The equations \eqref{mainequations}
apply to more than stationary configurations; they completely capture time 
dependent black hole dynamics as well. In this section we apply our equations
to a non stationary problem. We determine the most general solution to 
the equations \eqref{mainequations} in linearization about the Schwarzschild 
black hole solution. The solutions we find should map to the most general 
regular linearized solution of general relativity (with frequencies of order 
unity ) about the Schwarzschild black hole solution, i.e. the `light' 
quasinormal modes determined by EST (see the introduction). Indeed we find 
that the spectrum of the solutions we obtain agrees precisely with the 
spectrum of the light quasinormal modes of EST to appropriate order in 
$\frac{1}{d}$. We regard this match as a significant check on our membrane
equations \eqref{mainequations}

In the previous section we demonstrated that the membrane surface 
\begin{equation}\label{memsurf}
S^2+ r^2=1
\end{equation}
(recall $r^2= \sum_{i=1}^{p+1} w^i w_i$ where the index $i$ runs only over spatial 
coordinates) together with the velocity one form field on the surface of 
the membrane 
\begin{equation}
u=\left(-\frac{1}{S} \right)dt+\left(-\frac{r}{S}\right)dr
\end{equation}
solves the membrane equations \eqref{mainequations}, and that this solution 
corresponds to a Schwarzschild black hole of unit radius. In order to study 
small fluctuations about this solution we let the membrane fluctuate, i.e 
we let our membrane surface be given by 
\begin{equation} \label{smflum}
S-\sqrt{1-r^2}-a f(r,t,\theta^i)=0
\end{equation} 
Equivalently, the $S$ coordinate on the membrane is determined in terms of 
$r, \theta, v$ by the equation
\begin{equation} \label{sdef}
S=\sqrt{1-r^2}+a f(r,t,\theta^i)
\end{equation}
We also allow the velocity field to fluctuate, i.e we set
\begin{equation} \label{smfluv}
u=\left(-\frac{1}{S}+a \delta u_t\right)dt+\left(-\frac{r}{S}+a \delta u_r\right)dr+a \delta u_id\theta^i
\end{equation}
where $S$ is given by \eqref{sdef}. 
The parameter $a$ in \eqref{smflum} and \eqref{smfluv} measures the amplitude 
of the perturbation; through this section we work at linear order in $a$.

Our membrane effectively propagates in the $p+3$ dimensional space with 
metric 
\begin{equation} \label{metfs} 
ds^2= -dt^2+ dw_i dw^i + d S^2 = -dt^2 + dr^2 + r^2 d \Omega_{p}^2 + dS^2
\end{equation}

The normal to our perturbed surface is given, to ${\cal O}(a)$, by 
\begin{eqnarray} \label{nvf}
n&=&(\sqrt{1-r^2}+a r (1-r^2) \partial_rf)dS-a\sqrt{1-r^2}\partial_tfdt+(r-a\sqrt{(1-r^2)^3}\partial_r f)dr\nonumber\\
&&-a\sqrt{1-r^2}\partial_if d\theta^i
\end{eqnarray}
Once again the variable $S$ in \eqref{nvf} is given by \eqref{sdef}. 
Here $\theta^i$ 
are an arbitrary set of coordinates on the unit $p$ sphere in the 
metric \eqref{metfs}, and we have normalized $n$ to ensure that $n.n=1$ 
upto corrections of order $a^2$. 

Recall that the velocity field $u$ is constrained to obey $u.u=-1$ and 
$u.n=n_S-\frac{n^2}{n_S}$. These conditions allow us to solve for 
$\delta u_r$ and $\delta u_t$ in terms of $f$; we find  
\begin{eqnarray} \label{uif}
\delta u_r&=& \frac{a}{r(-1+r^2)}\left(r^2 f+(-1+r^2)\left(\partial_tf+r\partial_rf\right)\right) \nonumber \\
\delta u_t&=& r \delta u_r-a f 
\end{eqnarray}
It follows that we have $p+1$ independent fluctuation fields ($f$ and $u_i$).  

\subsection{The scalar equation}

We find it convenient to use $t, r, \theta^i$ as coordinates on the membrane. 
The dynamics of $f$ is controlled by the scalar equation 
(the first of \eqref{mainequations}). Evaluating the extrinsic curvatures 
associated with the normal vector field \eqref{nvf} (we performed the 
relevant computations on Mathematica), we find that $f$ obeys a linear 
partial differential equation of second order in derivatives in $t$ and $r$. 
The equation has no derivatives in the angular coordinates $\theta^i$.

It is useful to expand the fluctuation function $f$ as 
\begin{equation}\label{fexp}
f(r,t,\theta^i)=\sum_{lm} a_{lm}e^{-i w_{lm} t}G_{lm}(r) Y_{lm}(\theta^i)
\end{equation}
Here $l$ and $m$ refer to labels for scalar spherical harmonics of $SO(p+1)$. 
$l$ is a label for the  representation of $SO(p+1)$ \footnote{If we measure 
highest weights of representations by eigenvalues under rotations of 
independent two planes, the representation labelled by $l$ has highest 
weights $(l, 0, 0 \ldots 0)$.}  and $m$ is a collective label for the 
quantum numbers of states within a given representation. 

Plugging this expansion into the linear equation for $f$ yields an ordinary 
differential equation of $G_{lm}(r)$. Because the equation for $f$ had no 
derivatives in the angular directions, all functions $G_{lm}$ obey the same
equation. Making the replacement $G_{lm} \rightarrow G$ and 
$\omega_{lm} \rightarrow \omega$ we find that the equation takes the form
\begin{eqnarray} \label{defff}
&&-(r^4w^2+w(-2 i +w)+r^2(1+2 iw-2w^2))G(r)\nonumber\\
&&+r(-1+r^2)(2(r^2+iw-ir^2w)G'(r)+r(-1+r^2)G''(r))=0
\end{eqnarray}
where $'$ denotes derivative w.r.t. $r$. The two linearly independent solutions
to the equation \eqref{defff} are given by 
\begin{equation}\label{G}
G(r)=\frac{1}{\sqrt{i(-1+r^2)(i+4 w)}}\left(C_1 r^{\frac{1}{2}(1-\sqrt{1-4iw}+2iw)}+C_2r^{\frac{1}{2}(1+\sqrt{1-4iw}+2iw)}\right)
\end{equation}
Regularity of these solutions at $r=0$ requires the power of $r$ in
\eqref{G} to be a non negative integer. One of the two solutions above 
is proportional to $r^j$ \footnote{Which one depends on our precise definition
of the square root.} provided
\begin{equation}\label{w}
w_S(j)=-i(j-1)\pm\sqrt{j-1}
\end{equation}
While $j$ is required to be a non negative integer this condition 
is in itself not sufficient to ensure regularity for several different 
reasons. 

Let us first examine the lowest allowed value of 
$j$, namely $j=0$. At this value of $j$ the two frequencies \eqref{w} are 
$w_S(0)=0$ and $w_S(0)=2 i$. It is easily verified from \eqref{uif} 
that the mode with $w_s(0)=0$ has $\delta u_r=0$ and $\delta u_t$ 
regular everywhere. Consequently this mode is a regular small 
fluctuation of our membrane equations. Its physical interpretation is simple; 
it corresponds to rescaling the black hole. When, on the other hand, 
$w_s(0)=2i$, we find from \eqref{uif} that 
$\delta u_r \propto \frac{2}{r \sqrt{1-r^2}}$. \footnote{It is easily 
verified that for $j \geq 1$ $\delta u_r$ is well behaved at $r=0$. }
It follows that the velocity field corresponding 
to this mode is irregular at $r=0$, so this mode is irregular and so 
unacceptable. This is reassuring as $w=2i$ would have represented 
an instability, contradicting the well established stability of Schwarzschild 
black holes to small fluctuations in all dimensions. 
In summary, the only allowed frequency at $j=0$ is $w_S=0$.

Let us next turn to $j=1$. At this value of $j$ the quadratic equation 
for $w_S(1)$ yields a double zero; both roots in \eqref{w} are $w=0$. 
The fact that we find a double root indicates that $j=1$ is a special 
case that has to be dealt with separately; the two allowed time dependences
at this value of $j$ are actually $f$ constant in time and $f \propto t$
\footnote{Note that $t=\ln e^{t}$; the linear dependence in time 
follows from the application of a standard rule.  Recall that 
we plug ansatz $x^\alpha$ into a second order linear differential equation 
and find a double root for $\alpha$ at $\alpha=\alpha_c$ then a basis for
the solution space of the equation is given by  $x^\alpha_c$ and 
$x^\alpha_c \ln x $. }
These two `zero' modes have a simple physical interpretation. The constant
mode corresponds to an infinitesimal translation of the black hole, while 
the mode linear in time corresponds to an infinitesimal boost for the black 
hole. 

The next issue concerns the angular dependence of modes. Recall that 
the angular variation of $f$ did not enter our equation, so the reader may 
at first suspect that we have an infinite degeneracy of modes at every 
$j$, corresponding to the infinite number of possible angular variations, 
seemingly unconnected with radial behaviour. This is not the case. 
A scalar function of the form $r^j Y_{lm}(\theta)$
is regular at $r=0$ only if \footnote{The underlying reason for this is that 
spherical polar coordinates are  
singular at $r=0$.}
$$j = l,~ l+2, ~l+4  \ldots .$$ 
\footnote{This is most easily seen 
by noting that an $SO(p+1)$ spherical harmonic of degree $l$ is the 
restriction of a polynomial of degree $l$ (in the $p+2$ embedding Cartesian
variables) to the unit sphere.} It follows that we find a regular solution 
for $G_{lm}(r)$ provided
\begin{equation}\label{w1}
w=w_S(j), ~~~j = l, ~l+2, ~l+4 \ldots
\end{equation} 
In particular the minimum frequency for $G_{lm}(r)$ is $w_S(l)$. 

In summary the spectrum of frequencies of oscillation in the scalar 
sector is given by $\omega_S(j)$ for $j=0,1, 2, 3, \ldots $. For any 
given value of $j$, the modes that have frequency $\omega_S(j)$  
are scalar spherical harmonics  with 
$$l =j, j-2, j-4 \ldots $$
In other words modes with different angular momenta share the same frequency. 
Degeneracies are unusual in physics; when they occur they are usually 
explained by a symmetry. In the current context the relevant symmetry 
is obvious; fluctuations about the Schwarzschild black hole 
appear in representations of $SO(p+d+2)$. The different 
$SO(p+1)$ spherical harmonics that descend from a single $SO(p+d+2)$ 
spherical harmonic must have the same frequency. 

Now consider an $SO(p+1 +d+1)$ spherical harmonic that transforms in the 
representation $(j, 0, 0 \ldots 0)$ (as usual we label representations by 
the highest weights under rotations in the independent embedding two planes). 
This representation can be decomposed into a sum over irreducible representations
of $SO(p+1) \times SO(d+1)$. The number of representations that appear in this 
decomposition is very large. If we are interested only in those 
representations that happen to be singlets under $SO(d+1)$, it is not 
difficult to convince oneself that the branching rule is 
$$(j, 0, 0 \ldots 0) \rightarrow \sum_{m=0}^{[\frac{j}{2}]}
(j-2m, 0, 0 \ldots 0) \otimes I $$
In other words the set of $SO(p+1)$ spherical harmonics that have 
frequencies $\omega_j$ are precisely obtained from a single angular momentum 
$j$ spherical harmonic of $SO(p+d+2)$ subject to the constraint that we 
restrict attention only to $SO(d+1)$ singlets. 

In summary, the spectrum of scalar fluctuations is given by $\omega_S(j)$ for 
non negative $j$. Eigenmodes with frequency $\omega_S(j)$ are given by 
the projection onto $SO(d+1)$ singlets of an angular momentum $j$ spherical 
harmonic of $SO(d+p+2)$. This result is in 
perfect agreement with the frequency of light scalar quasinormal modes about the 
Schwarzschild black hole obtained by EST in Equation $(5.29)$  
of \cite{Emparan:2014aba}. 

\subsection{The vector equation}
 
In the parametrisation of this subsection the vector equations of 
\eqref{mainequations} become
\begin{equation}\label{veceqpert}
(-2+r^2)u_i+r(1-r^2)\partial_r u_i+(1-r^2)\partial_tu_i=-(1-r^2)\partial_i f
\end{equation} 
where $u_i$ is a one form field that lives on the unit $p$ sphere.

\eqref{veceqpert} is an inhomogeneous or  forced linear differential 
equation in the one form variable $u_i$. The solutions of any such equation 
take the form $H+ P$ where $P$ is a particular solution and $H$ is the 
general solution to the homogeneous part of the equation (i.e. \eqref{veceqpert}
with the RHS set to zero). 
 
The particular solution may easily be seen 
to be regular (the key point here is that the frequencies of the homogeneous 
part of \eqref{veceqpert}, which we work out below, never coincide with 
the scalar frequencies of the previous subsection) and is of no 
particular interest to us; we do not bother to work it out. 
\footnote{It is natural to choose the particular solution of this equation 
as a linear combination of terms with the scalar frequencies 
listed in \eqref{w}. At any given frequency, the particular (or forced) 
solution is proportional to the coefficient of the scalar mode in $f$ at 
the same frequency, and multiplies a regular function of $r$.} Our interest
is in the spectrum of the homogeneous solutions, to which we now turn. 

The most general solution to the homogeneous equation
\begin{eqnarray}\label{pertequations}
(-2+r^2)u_i+r(1-r^2)\partial_r u_i+(1-r^2)\partial_tu_i=0
\end{eqnarray}
is easily determined. 
It is given by a linear combination of terms of the form 
$$u^\omega_i(r)= e^{-i \omega t} r^{2+i \omega} q_i(\theta) $$
Here $q_i(\theta)$ is an arbitrary one form field on the unit sphere. 

The solutions above are clearly irregular at $r=0$ unless $2-i\omega$ is
a non negative integer. This is a necessary and not a sufficient condition 
for regularity. 

In order to understand the sufficient conditions for regularity
more clearly it is useful to decompose 
\begin{equation}\label{vdecom}
q_i=v_i+\partial_i\lambda,\quad \mbox{where}~~ \nabla_iv^i=0
\end{equation}
and $\lambda$ is a scalar function on the sphere. 
$v_i$ may then may be decomposed in a basis of vector spherical harmonics, 
while $\lambda$ may be decomposed in a basis of scalar spherical harmonics. 

Let us define the $l^{th}$ vector spherical harmonic as the set of vector 
functions on the sphere that transform in the $(l, 1, 0, \ldots 0)$
representation of $SO(p+1)$ (we label representations by eigenvalues of the 
rotation operators in the independent two planes in the embedding $p+1$ 
dimensions). It is not difficult to convince oneself that a one form field 
whose angular part is proportional to the $l_V^{th}$ vector spherical harmonic
is regular if and only if the function that multiplies it is a linear combination
of terms of the form $r^j$ where 
$$j=l_V+1, l_V+3, l_V+5 \ldots .$$

Let us now turn to the regularity of $d \lambda$. As we have already noted 
in the last subsection, a scalar function $\lambda$ is regular if and only if 
the coefficient of the $l_S^{th}$ scalar spherical harmonic (defined as the set 
of functions in the representation $(l_s, 0, 0 \ldots , 0)$ of $SO(p+1)$) 
is a linear sum of terms of the form $r^j$ where 
$$j=l_S, l_S+2, l_S+4 \ldots $$
Notice that for $l_S=0$ $d \lambda=0$ so that the minimum value for $l_S$ is 
unity. 

Putting all these facts together, we find that we have regular solutions 
to the homogeneous vector equation at 
\begin{equation}\label{omev}
\omega=\omega_V(j)= -i(j-1), ~~~j=1, 2 \ldots  
\end{equation}
This spectrum is in perfect agreement with 
Equation (5.21) of \cite{Emparan:2014aba}. \footnote{We can also list the eigen functions.  
The allowed values of $j$ for the frequencies of the $l_v^{th}$ vector 
spherical harmonic are $j=l_V, l_V+2, l_V+4, \ldots$ while 
the allowed $j$ for the frequencies for the gradient 
of the $l_S^{th}$ scalar spherical harmonic are 
$j=l_s-1, l_s+1, l_s+ 3 \ldots .$. We 
several degeneracies in the frequencies at any particular value of $j$. 
As in the previous subsection we presume that all representations with a given $\omega_V(j)$ 
have their origin in a single vector spherical harmonic representation of $SO(d+p+2)$, but 
we have not carefully verified this claim. }

Note that $w_V$ vanishes at $j=1$. This mode corresponds to turning on 
an infinitesimal rotation of the black hole (recall that rotations, 
like angular momenta, transform in the representation $(1,1, 0 \ldots 0)$  
- i.e. the lowest vector spherical harmonic representation - or the 
appropriate orthogonal groups. All modes with $j \geq 2$ have an imaginary 
part to their frequency that corresponds to an exponential decay in time. 
It follows that the Schwarzschild black hole at large $D$ is stable to 
vector fluctuations (as it was to scalar fluctuations) in accordance 
with well known results.

\vskip .8cm
\section*{Acknowledgements}
We would like to thank P. Ajit, B. Craps, Y. Dandekar, S. Das, R. Emparan, 
A. Ghosh, R. Gopakumar 
V. Hubeney, K. Imbasekar, S. Mandal, M. Mandlik, S. Nadkarni, P. Nayak, M. Rangamani, A. Sen, N. Seiberg, S. Trivedi, S. Wadia and A. Yarom for useful 
discussions.  The work of S.B. was supported by an India Israel (ISF/UGC) joint research 
grant. The work of A.D is supported by an Inspire Fellowship from DST, India.
S.M. would like to acknowledge the hospitality of ICTS where this work  
was intitiated and of the Harish Chandra Research Insitute, IISER Kolkata, 
the University of Swansea and the Galileo Galilei Institute while this work 
was in progress. The work of S.M. was supported by an India Israel (ISF/UGC) 
joint research grant. We would all also like to acknowledge our
debt to the people of India for their steady and generous support to 
research in the basic sciences.

\appendix

\section{Technical details}

\subsection{Simplification of the dimensionally reduced equations} \label{dr}

Upon varying \eqref{redlagrangian} w.r.t the scalar $\phi$ we find the equation of motion 
\begin{equation}\label{phvar}
 R+ e^{-\phi}(d-2)(d-1) - \frac{d(d-1)}{4} (\partial \phi)^2 -(d-1)\nabla^2 \phi=0
\end{equation}
Varying w.r.t. $g^{\mu\nu}$ we find
\begin{equation}\label{varlag}
 R_{\mu\nu}-\frac{R}{2} g_{\mu\nu} = \frac{d}{2} \left( \nabla_\mu \nabla_\nu \phi- \nabla^2 \phi g_{\mu\nu} \right) + \frac{d(d-1)}{2} e^{-\phi} g_{\mu\nu} 
 +\frac{d}{4} \nabla_\mu \phi \nabla_\nu \phi - \frac{d(d+1)}{8} (\nabla \phi)^2 g_{\mu \nu}
\end{equation}
\eqref{varlag} may be somewhat simplified by subtracting half of 
\eqref{phvar} multiplied by $g_{\mu\nu}$ and the equations of motion may more simply be written as 
\begin{equation}\label{reom}\begin{split}
&R+ e^{-\phi}(d-2)(d-1) - \frac{d(d-1)}{4} (\partial \phi)^2 -(d-1)\nabla^2 \phi
=0\\
&R_{\mu\nu}= \frac{d}{2} \nabla_\mu \nabla_\nu \phi +\frac{d}{4} \nabla_\mu \phi \nabla_\nu \phi + g_{\mu\nu} \left( (d-1) e^{- \phi} - 
\frac{d}{4} (\partial \phi)^2 - \frac{1}{2}{\nabla^2} \phi \right)  \\
\end{split}
\end{equation}
By subtracting the trace of the second equation from the first we find 
equivalently
\begin{equation}\label{rreom}\begin{split}
& e^{-\phi}(d-1) - \frac{d}{4} (\partial \phi)^2 - \frac{1}{2}\nabla^2 \phi
+ \frac{p+2}{d-2}\left( (d-1) e^{- \phi} - 
\frac{d}{4} (\partial \phi)^2 - \frac{1}{2} {\nabla^2} \phi \right)  =0\\
&R_{\mu\nu}= \frac{d}{2} \nabla_\mu \nabla_\nu \phi +\frac{d}{4} \nabla_\mu \phi \nabla_\nu \phi +  g_{\mu\nu} \left( (d-1) e^{- \phi} - 
\frac{d}{4} (\partial \phi)^2 - \frac{1}{2}{\nabla^2 \phi} \right) \\
\end{split}
\end{equation}
where the indices $\mu, \nu$ run over $p+3$ values. Notice that the two 
terms in the scalar equation are simply proportional to each other, so that 
the scalar equation is equivalent to 
$$\left( (d-1) e^{- \phi} - 
\frac{d}{4} (\partial \phi)^2 - \frac{1}{2}{\nabla^2} \phi \right)=0.$$
Given this condition, the second term in the vector equation vanishes, and 
our system of equations reduces to 
\begin{equation}\label{rrreoma}\begin{split}
& e^{-\phi}(d-1) - \frac{d}{4} (\partial \phi)^2 - \frac{1}{2}\nabla^2 \phi
 =0\\
&R_{\mu\nu}= \frac{d}{2} \nabla_\mu \nabla_\nu \phi +\frac{d}{4} \nabla_\mu \phi \nabla_\nu \phi  \\
\end{split}
\end{equation}

\subsection{Details of black branes} \label{bb}

\subsubsection{Standard coordinates for the black brane}

In order to verify that \eqref{bbf} is indeed a solution, it is useful to 
move to a convenient coordinate system. Let $x_0$ denote the (constant) 
radius of the $d$ sphere about the point of interest so that  
$$\frac{1}{x_0^2}=e^{-\phi}=\frac{\chi^2}{4}$$ 
We then define the coordinate $X$ by the requirement
$$2 \frac{dX}{x_0}= \chi - 2 dR$$
We also define the coordinate $V$ by the requirement that 
$$dV=O$$
These definitions are sensible because the components of $\chi$ and $O$
are constants
\footnote{As emphasized in the previous footnote, this definition fixes 
$V$ only upto a scaling. We will fix the ambiguity of the scale factor below.}
At this stage we have three special coordinates; $R$, $X$ and $V$. 
Let the remaining $p$ coordinates be denoted by $y^i$.  
We employ coordinate redefinitions of the form 
$y^i \rightarrow y^i + C^i_R R + C^i_V V + C^i_X X$ to remove all cross terms 
between $y^i$ and $(X, V, R)$. We then employ redefinitions of the form 
$Y^i= \zeta^i_j y^j$  to set the metric in the $Y^i$ directions 
to $\delta_{ij}$. With all these conventions we now proceed to display 
an explicit form of the metric \eqref{bbf}.

The choices we have made in the previous paragraph clearly ensure that 
the metric is a direct sum of $dY^i dY^i$ and a three dimensional metric 
involving $dX$, $dV$, and $dR$. We will now determine the form of this 
three dimensional metric. 

The conditions \eqref{conditions} ensure that there are no cross 
terms between 
$dX$ and either $dV$ or $dR$. It follows that the metric takes the form 
$$ \frac{1}{1-a x_0^2} dX^2 + b dR^2  + 2 C dR dV - D dV^2$$
where $a$, $b$, $C$ and $D$ are arbitrary constants. 
The fact that $dV$ is null sets $b$ to zero. As we have emphasized in the 
footnotes above, the coordinates $V$ and $R$ are ambiguous upto a coordinated
rescaling and shift. We choose to fix this ambiguity by setting $C=1$. 
Making this choice, the condition on the squared norm of $\chi$ sets $D=a$, 
and the metric \eqref{bbf} takes the explicit form 
\begin{equation}\label{peaa} \begin{split}
&e^{-\phi}=\frac{1}{x_0^2}, ~~~\chi=2 d R + \frac{2}{x_0} dX \\
&ds^2= 2 dR dV - a \left(1-e^{-R} \right) dV^2 + dY^i dY^i + \frac{dX^2}{1-a x_0^2}
\\
\end{split}
\end{equation}
We have
used an additional shift in $R$ to scale the factor $a$ outside $e^{-R} dV^2$
to ensure that the event horizon lies at $R=0$.

\subsubsection{Uniform scaling of black branes}

The black brane metric \eqref{pe} appears in a two parameter family labelled
by $x_0$ and the parameter $a$. There is a simple explanation for the parameter
$a$; its existence follows from the fact that Einstein's equations in 
$p+d+3$ dimensions are invariant under an overall rescaling of the metric. In terms of the reduced fields described above, the map 
\begin{equation}\label{rescaling}
(g_{\mu\nu}, e^\phi) \rightarrow (A g_{\mu\nu}, Ae^{\phi})
\end{equation} 
takes solutions to solutions. It is easily verified that this rescaling operation 
does indeed map black branes to black branes. More concretely, the rescaling 
operation \eqref{rescaling} applied to the black brane \eqref{pe} yields a new 
black brane 
\begin{equation}\label{pe} \begin{split}
&e^{-\phi}=\frac{1}{{\tilde x}_0^2}, ~~~\chi=2 d R + \frac{2}{ {\tilde x}_0} 
d{\tilde X} \\
&ds^2= 2 dR d{\tilde V} - {\tilde a} \left(1-e^{-R} \right) d {\tilde V}^2 + 
G_{ij}dY^i dY^j + \frac{d {\tilde X}^2}{1-a {\tilde x}_0^2}
\\
\end{split}
\end{equation}
with 
\begin{equation}\label{redefs}
{\tilde a}= \frac{a}{A}, ~~~{\tilde x}_0^2=A x_0^2, ~~~{\tilde V}= AV, ~~~
{\tilde X}= \sqrt{A} X
\end{equation} 
The important point here is that this rescaling changes the value of $a$, and 
so generates black branes with all values of $a$ starting with black branes 
at $a=1$.

\subsection{Location of the event horizon}\label{eheh}

We will now demonstrate that the spacetime \eqref{fssa} has an event horizon 
located `near' to the surface $\psi=1$. As usual we consider a solution 
of Einstein's equations that eventually settles down into a rotating black hole
solution. In this context the event horizon is the unique null manifold 
that reduces, at large time, to the event horizon of the eventual black hole.
In order to compute the event horizon manifold of the spacetime 
\eqref{fssa}, let us set up a coordinate system in the neighbourhood of 
the membrane. Let $\psi$ be one of our coordinates and let $z^a$ define 
the remaining $p+2$ coordinates. Let the event horizon manifold be given by 
the equation 
\begin{equation}\label{eh}
\psi = \psi_H(z^a)
\end{equation}
We will now demonstrate that 
$$\psi_H= 1 + \frac{\psi_1}{d} + \ldots $$
The argument is very simple. The event horizon is determined by the requirement 
that the one form 
$$ d \psi + \partial_a \psi_H dz^a$$
is null. Let us work in a small patch around the point $x_0^\mu$ which lies on 
the horizon. We have argued above that the metric is that of the black 
brane with the role of $\psi$ played by the coordinate $R$. Ignoring 
derivatives the null surface lies at the location $e^{-R}=a$. While 
derivatives are nonzero, they are of order $\frac{1}{d}$. It follows that
the true null surface deviates from $e^{-R}=a$ only at order $\frac{1}{d}$. 
In other words we have not merely established our result; we have also 
determined the precise location of the event horizon at leading order in 
$\frac{1}{d}$: it is simply the event horizon of the local black brane.

\subsection{Equivalence in data off the membrane} \label{inequiv}

In this short appendix we argue that the replacement
$$O \rightarrow O+ \delta O, ~~~B \rightarrow \alpha B$$
where $\delta O$ vanishes on the membrane, and $\alpha$ is any function that 
does not blow up or vanish on the membrane, generates equivalent metrics
of the form \eqref{fssa}.

In order to see this we note that in the large $d$ limit 
$$\frac{1}{\psi^{p+d}} \sim e^{-d(\psi -1)}$$
As $\psi$ moves away from unity, in other words, this function dies off 
extremely rapidly. In other words the metric \eqref{fssa} only cares about 
the vector field $O$ in a neighbourhood of the membrane where $\psi-1=
{\cal O}(1/d)$. In this region the first Taylor expansion of $O$ - 
and the second Taylor expansion of $\psi$ - away from the membrane each 
contribute to the metric at order $\frac{1}{d}$. To leading order, in other 
words, the metric cares only about the location of the membrane, 
 $O$ on the membrane, and $d \psi$ on the membrane. 

We will now argue that $d\psi$ on the membrane is also completely determined 
by the shape of the membrane. $d \psi$ is proportional to the 
the normal vector field of the membrane surface. As a consequence it is 
fixed by the shape of the surface upto a position dependent normalization. 
Now the normalization of $\psi$ is unambiguously determined by 
the requirement $(S d\psi - d S). dS=0$. Indeed our definition of 
$\psi$ automatically obeys this equation. As the normalization of $\psi$ is 
fixed by a physical requirement, it follows that $d\psi$ is entirely 
determined by the shape of the membrane. 
\footnote{The fact that the normalization of $d\psi$ is physical is closely 
related to the following observation. $d \psi$ is proportional to $dB$. 
Under the rescaling $B \rightarrow \alpha B$, $dB \rightarrow \alpha dB$ on the 
membrane. However $d\psi$ is unaffected by this scaling on the membrane.}

We conclude that, to leading order, inequivalent ansatz are parametrized by 
the shape of the membrane and $O$ on the membrane.

\subsection{Details concerning Myers-Perry black holes} \label{mp}

The metric for a rotating black hole in arbitrary dimensions 
is well known \cite{Myers:1986un,Myers:2011yc}. 
The black hole metric is given in Eddington-Finkelstein coordinates by
in $2 N +2$ dimensions   
\begin{equation}\begin{split}\label{rotblacansatze}
ds^2 &= 2\left(dv-\sum_{i=1}^{N}a_i\mu_i^2d\theta_i\right)d\rho +\sum_{i=1}^{N}(\rho^2 + a_i^2)(d\mu_i^2+\mu_i^2d\theta_i^2)-dv^2\\ &+\left(\frac{m\rho}{\Pi F}\right)(dv-\sum_{i=1}^{N}a_i\mu_i^2d\theta_i)^2+\rho^2d\alpha^2,
\end{split}
\end{equation}
while in $2N+1$ dimensions the metric is given by 
\begin{equation}\begin{split}\label{rotblacansatzo}
ds^2 &= 2\left(dv-\sum_{i=1}^{N}a_i\mu_i^2d\theta_i\right)d\rho +\sum_{i=1}^{N}(\rho^2 + a_i^2)(d\mu_i^2+\mu_i^2d\theta_i^2)-dv^2\\ &+\left(\frac{m\rho}{\Pi F}\right)(dv-\sum_{i=1}^{N}a_i\mu_i^2d\theta_i)^2
\end{split}
\end{equation}
In both \eqref{rotblacansatze} and \eqref{rotblacansatzo}, the black hole 
is characterized by a mass parameter $m$ and 
the $N$ independent angular velocities $a_i$ in the 
$N$ angular directions $\theta_i$. Apart from $\theta_i$, the coordinates 
in both the metrics above consist of $\rho$, $v$ the $N$ `direction cosines' 
$\mu_i$ yielding of total of $2N+2$ coordinates in all. Of course this is one 
coordinate too many for the odd dimensional metric, in which case the coordinates
$\mu_i$ are constrained to obey 
\begin{equation}\label{contfo}
\sum_i^N\mu_i^2=1
\end{equation} 
In the case of the even dimensional black hole, the additional `coordinate'
$\alpha$ in  \eqref{rotblacansatze} is given in terms of direction cosines by 
\begin{equation}\label{contfe}
\alpha^2+ \sum_i^N\mu_i^2=1
\end{equation}  
In both metrics the quantities 
$\Pi$ and $F$ are defined by 
\begin{equation}\begin{split} \label{pifdef}
\Pi &= \prod_{i=1}^{N}(\rho^2+a_i^2),\\
F &= 1-\sum_{i=1}^{N}\frac{a_i^2\mu_i^2}{\rho^2+a_i^2}\\
\end{split}
\end{equation}

The rotating black holes can be recast in `flat space' type coordinates as 
follows. In the case of the even dimensional black hole we move to the new 
coordinates $x_i, y_i, z, t$ ($i=1 \ldots N$) defined by 
\begin{equation}\label{mpcoordtranse}
\begin{split}
x_i &= \mu_i(\rho\cos \theta_i - a_i\sin \theta_i),\\
y_i &= \mu_i(\rho\sin \theta_i + a_i \cos \theta_i),\\
z &= \rho\alpha,\\
t& = v-\rho
\end{split}
\end{equation}
In the case of the odd dimensional black hole we use use the coordinates 
$x_i$, $y_i$ and $t$ defined in a similar manner by 
\begin{equation}\label{mpcoordtranso}
\begin{split}
x_i &= \mu_i(\rho\cos \theta_i - a_i\sin \theta_i),\\
y_i &= \mu_i(\rho\sin \theta_i + a_i \cos \theta_i),\\
t&=v-\rho
\end{split}
\end{equation}
In terms of the new coordinates the even dimensional black hole 
\ref{rotblacansatze} takes the form 
\begin{equation}\label{rotblackansatxef}
\begin{split}
ds^2 &= -dt^2 + \sum_{i=1}^{N}(dx_i^2 + dy_i^2) + dz^2 \\&+ \frac{m\rho}{\Pi F}\left(dt+\left(1-\sum_{i=1}^{N}\frac{\mu_i^2a_i^2}{\rho^2+a_i^2}\right)d\rho-\sum_{i=1}^{N}\frac{a_i}{\rho^2+a_i^2}(x_idy_i-y_idx_i)\right)^2.
\end{split}
\end{equation}
whereas the odd dimensional black hole \ref{rotblacansatzo} turns into  
\begin{equation}\label{rotblackansatxof}
\begin{split}
ds^2 &= -dt^2 + \sum_{i=1}^{N}(dx_i^2 + dy_i^2)  \\&+ \frac{m\rho}{\Pi F}\left(dt+\left(1-\sum_{i=1}^{N}\frac{\mu_i^2a_i^2}{\rho^2+a_i^2}\right)d\rho-\sum_{i=1}^{N}\frac{a_i}{\rho^2+a_i^2}(x_idy_i-y_idx_i)\right)^2.
\end{split}
\end{equation}
The quantities $\mu_i$ and $\rho$ in \eqref{rotblackansatxef}  and 
\eqref{rotblackansatxof} are now functions of our new coordinates. 
In the even dimensional case these qantities are determined by by the 
equations 
\begin{equation}\label{rhoconste} \begin{split}
& \mu_i^2=\frac{x_i^2+y_i^2}{\rho^2+a_i^2} \\
& \sum_{i=1}^N \frac{x_i^2+y_i^2}{\rho^2+a_i^2} + \frac{z^2}{\rho^2} = 1
\end{split}
\end{equation}
while in the odd dimensional case the relevant relations are  
\begin{equation}\label{rhoconsto} \begin{split}
& \mu_i^2=\frac{x_i^2+y_i^2}{\rho^2+a_i^2} \\
& \sum_i^N\frac{x_i^2+y_i^2}{\rho^2+a_i^2} =1\\
\end{split}
\end{equation}

In this paper we are interested in the special case of rotating black holes 
with non-zero rotations turned on in only $q$ two planes with $q \ll D$. 
Given such a black hole (as mentioned in the main text) we choose $p=2q-1$. 
Let us now change notation so that $x_i$ and $y_i$ are coordinates of the 
two planes with non-zero rotations, i.e. $i=1 \ldots q$. We group the other 
$x_i, y_i$ coordinates (the coordinates of the remaining two planes in which 
the angular velocities all vanish) together as $z_M$ coordinates. In 
the even dimensional context \eqref{rotblackansatxef} 
these $z_M$ coordinates appear on equal footing with the coordinate $z$ and we 
group that coordinate together with $z_M$. 
With these definitions the index $M$ runs over $d+1$ values and the black hole 
metric, in both odd and even dimensions, takes the form 
\begin{equation}
\begin{split}
ds^2 &= -dt^2 + \sum_{i=1}^{q}(dx_i^2 + dy_i^2) +\sum_{M} dz_M^2 \\
&+ \frac{m\rho}{\Pi F}\left(dt+\left(1-\sum_{i=1}^{q}\frac{\mu_i^2a_i^2}{\rho^2+a_i^2}\right)d\rho-\sum_{i=1}^{q}\frac{a_i}{\rho^2+a_i^2}(x_idy_i-y_idx_i)\right)^2.
\end{split}
\end{equation}
where $\mu_i^2= \frac{x_i^2+y_i^2}{\rho^2+a_i^2}$ as above, and the equation 
that implicitly determines $\rho$ is 
\begin{equation}\label{rdimp}
\sum_{i=1}^{[\frac{p+1}{2}]} \frac{x_i^2+y_i^2}{\rho^2+ a_i^2} 
+ \frac{S^2}{\rho^2}=1
\end{equation}
where $S^2= \sum_{M=1}^{d+1} z_M^2$ as in the main text.

\section{Details of the perturbative procedure}\label{method}

We move to a patch centred around $x_0^\mu$ and use the following patch 
coordinates
\begin{equation}\label{scale_coord}
\begin{split}
&\frac{R }{d}= (\psi-1),~~\frac{V}{d} = e^{h(x_0)} O_\mu(x_0)(x^\mu - x^\mu_0),~~\frac{Y}{d}= S  -s_0~\psi\\
&\frac{y^i }{d}= {\cal P}^i_\mu(x_0)(x^\mu-x^\mu_0)
\end{split}
\end{equation}
In these coordinates the rescaled  zeroth order metric $G_{a b}$ 
and the one form $\chi$ take the following form.
\begin{equation}\label{metriczero}
\begin{split}
ds^2= &~2 s_0 e^{-h_0} dR ~dV -e^{-2h_0}F(R)dV^2\\
&  +\frac{dY^2}{c^2} + \sum_{i=1}^p dy^i dy^i + {\cal O}\left(\frac{1}{d}\right)\\
\chi=&\frac{2dY }{s_0}+ 2 dR+ {\cal O}\left(\frac{1}{d}\right),~~e^{-\phi} = \frac{1}{s_0^2} + {\cal O}\left(\frac{1}{d}\right)
\end{split}
\end{equation}
where $$F(R)=\bigg((1-c^2)- e^{2h_0-R}\bigg)$$
Ignoring corrections of order $\frac{1}{d}$, \eqref{metriczero} is the 
black brane metric. \footnote{\eqref{metriczero} can 
be cast into the standard form \eqref{pea} by the rescaling 
$V =\left(\frac{e^{h_0}}{s_0}\right) \tilde V$ and shifting $R=\tilde R+[2h_0 -\log(1-c^2)]$. The `$a$' parameter of 
the standard form metric may be verified to be given by 
$$a = \frac{1-c^2}{s_0^2}$$
In our computations below, however, we do not perform this scaling and shift; we 
work with the $V$ rather than the  $\tilde V$ coordinate.}

In these coordinates the terms listed in the second line onwards of \eqref{prathomic} 
takes the form 
\begin{equation}\label{metricone}
\begin{split}
ds^2|_{{\cal O}\left(\frac{1}{d}\right)}&=\frac{e^{-R}}{d}\bigg[K_1(R) dV^2 + 2 K_2(R)dV ~dY+ K_3(R) dY^2\\
&~~~~~~~ + 2K_4(R) dR~ dY+ K_5(R) dy^i dy^i\\
&~~~~~~~+2Q^i_1(R)dV~ dy^i + 2Q^i_1(R)dY~ dy^i + {\cal T}_{ij}(R)dy^i dy^j\bigg]
\end{split}
\end{equation}
As we have explained in the main text, at ${\cal O}\left(\frac{1}{d}\right)$ 
the scalar, vector and the tensor data are effectively constants (equal to their values at the point $x^\mu_0$) as derivatives of these expressions 
contribute to the metric only at ${\cal O}\left(\frac{1}{d}\right)^2$ and 
higher. It follows that  
\begin{equation*}
\begin{split}
&K_a(R) = \sum_{A=0}^7 K_a^A(R) S_A(x^\mu_0)\\
&Q^i_a(R) = \sum_{A=1}^5Q^A_a(R)V^i_A(x^\mu_0)\\
&{\cal T}_{ij}(R) = \sum_{A=1}^2{\cal T}_A(R) T^{ij}_A(x^\mu_0)
\end{split}
\end{equation*}
 Substituting \eqref{metricone} in Einstein equation and the scalar field 
equation as given in \eqref{redeom} we obtain ordinary differential 
equations for various metric functions. $SO(p)$ symmetry ensures that 
the equations for the scalar functions $K_a$, vector functions $Q^i_a(R)$ 
and tensor functions ${\cal T}_{ij}(R)$ are mutually decoupled. Of course the 
equations for all functions of a given symmetry (e.g. scalars) all mix. 
Nonetheless it turns out that the equations are rather simple and explicitly 
solvable \footnote{As the equations for the unknowns are all linear it is 
possible in principle- and useful in practice - to obtain solutions for 
pieces of one derivative background data turned on one at a time and then 
to add the results.}

Recall that our data included an antisymmetric tensor; as there is no
antisymmetric tensor in the metric, it follows even without a calculation that 
there is no term in the correction metric proportional to this piece of data.

\subsection{Scalar sector}
In this sector the relevant equations are $E_{RR},~E_{RV},~E_{RY},~E_{VV},~E_{VY},~E_{YY}$ and $E_{tr}\equiv\sum_{i}E_{ii}$. Here by $E_{\mu\nu}$ we denote the second equation in \eqref{redeom}.

In order to solve these equations we found it useful to first study 
a linear combination of  $E_{RV}$ and $E_{VV}$  which turns out to give 
a decoupled equation for $K_2(R)$.
\begin{equation}\label{eqn1}
\begin{split}
E_1 \equiv &~ s_0 e^{h_0+R}E_{VV} +e^R F(R)E_{RV} \\
=&~\sum_{A=0}^7S_A\bigg[\frac{c^2e^{R}}{2s_0}F(R)^2 \left[\frac{e^{-R}K_2^A(R)}{F(R)}\right]' + {\mathfrak S_2^A}(R)\bigg]
\end{split}
\end{equation}
where $'$ denotes the derivative w.r.t. $R$.
The linear combination of equations presented in the first line \eqref{eqn1} 
evaluates (by explicit computation) to the second line of that equation. 
The first term in \eqref{eqn1} is homogeneous in the unknown $K_2^A(R)$
while the second term,  ${\mathfrak S}_2^A(R)$, is the source term proportional 
to the scalar data $S^A$. The explicit expressions for these source terms 
are
\begin{equation}\label{sourceS2}
\begin{split}
{\mathfrak S}_2^0(R)&=-\frac{c^2e^{h_0}}{2s_0}R\\
{\mathfrak S}_2^1(R)&=e^{h_0}\left(-\frac{c^6-6 c^4+8 c^2-1}{2 c^2
   \left(c^2-1\right)}+\frac{\left(c^2-2\right) R}{2
   \left(c^2-1\right)}-\frac{\left(c^2-2\right) e^{2h_0 -R}}{2
   \left(c^2-1\right)}\right)\\
 {\mathfrak S}_2^2(R)&=  \frac{e^{h_0}}{2} \left(2
   c^2+\frac{1}{c^2}-6-R+\frac{\left(2
   c^2-3\right) e^{2
   h_0-R}}{c^2-1}\right)\\
   {\mathfrak S}_2^3(R)&= \frac{e^{h_0}}{2}\left(1-c^2-e^{2h_0 -R}\right)\\
    {\mathfrak S}_2^4(R)&=\frac{e^{h_0}}{2},~~ ~~~{\mathfrak S}_2^5(R)=e^{h_0}(-2 + c^2)\\
     {\mathfrak S}_2^6(R)&=-2 {\mathfrak S}_2^7(R)= (1-c^2)e^{h_0}\\
\end{split}
\end{equation}
\eqref{eqn1} is easily solved by integration and we obtain 
\begin{equation}\label{samadhan1}
\begin{split}
K_2^A(R)=\frac{2 s_0e^{R}F(R)}{c^2}\int_R^\infty d\rho\left(\frac{e^{-\rho}{\mathfrak S}_2^A(\rho)}{F(\rho)^2}\right)
\end{split}
\end{equation}
The integration constant is fixed using the boundary condition that for large $R$, $K_2(R)$ should behave at most like a polynomial in $R$.

Now the function $F(R)$ has a zero at $R_0=2h_0 -\log(1-c^2)$. It follows that 
the integrand in \eqref{samadhan1} can be expanded around $\rho=R_0$ as 
$$\left(\frac{e^{-\rho}{\mathfrak S}_2^A(\rho)}{F(\rho)^2}\right)
=\frac{B^A_1}{(\rho - R_0)^2}+ \frac{B^A_2}{\rho-R_0} + {\rm regular}$$
Upon integrating, it follows that 
$$K_2(R) \propto 
(R-R_0) \sum_{A}\left( \frac{B_1^A S_A}{R-R_0} + {B_2^A S_A} \ln(R-R_0) + {\rm regular} 
\right) $$
While the terms proportional to $B_1^A$ above are perfectly analytic, 
the terms proportional to $B_2^A$ are potentially singular (non analytic). 
It follows that $K_2(R)$ solutions is regular everywhere outside its event horizon (as we demand) if and only if
$$ \sum_{A} B_2^A {S}_A=0.$$
This equation may be rewritten as 
\begin{equation}\label{absteqn}
\text{Equation of motion} \propto \sum_{A=0}^7 S^A\left(\partial_R{\mathfrak S}_2^A(R)|_{R=R_0}\right) =0
\end{equation}
upon explicitly evaluating the derivatives in \eqref{absteqn} we find the 
scalar equation in \eqref{mainequations}.
After imposing the equation of motion $K_2(R)$ is perfectly regular.

Next we solve $E_{RY}=0$. If we substitute the solution for $K_2(R)$ and the equation of motion, $E_{RY}$ becomes very simple: 
\begin{equation}\label{ery}
E_{RY}\equiv \frac{c^2}{2s_0}\partial_R\left(e^{-R}K_3(R)\right)=0
\end{equation}
The only normalizable solution to this equation $$K_3(R)=0$$

$E_{tr}$ gives another decoupled source free equation for $K_5(R)$.
\begin{equation}\label{etr}
E_{tr}\equiv\frac{e^{-R}}{2 s0^2}\partial_R\bigg(e^RF(R)~\partial_R\left(e^{-R}K_5(R)\right)\bigg)=0
\end{equation}
If we impose regularity at $R=R_0$ and normalizability at $R=\infty$
the only allowed solution is $$K_5(R)=0$$

After we substitute the solutions for $K_2(R),~K_3(R)$ and $K_5(R)$, the equation $E_{RR}$ becomes the simple source free equation for  $K_4(R)$.
\begin{equation}\label{err}
E_{RY}\equiv \frac{c^2}{2s_0}\partial_R\left(e^{-R}K_4(R)\right)=0
\end{equation}
From here it follows 
$$K_4(R)=0$$

Finally we solve for $K_1(R)$. We use the scalar field equation (i.e., 
the first equation in \eqref{redeom}) for this purpose. After substituting 
in the solution for all the other $K_i(R)$ as well as the equation of motion, 
this scalar equation takes the form 
\begin{equation}\label{escalar}
\begin{split}
E_\phi\equiv \sum_{A=0}^7S_A\left(\partial_R K_1^A(R)+\mathfrak{S}_1^A(R)\right)=0
\end{split}
\end{equation}
where 
\begin{equation}\label{sourceS1}
\begin{split}
{\mathfrak S}_1^0(R)&=\frac{-c^6+11 c^4-23
   c^2+11+\left(c^2-1\right)^2 \left(3 c^2-8\right) R}{\left(c^2-2\right)^2 \left(c^2-1\right)}\\
{\mathfrak S}_1^1(R)&=0\\
 {\mathfrak S}_1^2(R)&= \frac{2 s_0 \left(1-2 c^4+c^2 (R+3)\right)}{c^2 \left(c^4-3 c^2+2\right)}\\
   {\mathfrak S}_1^3(R)&=\frac{s_0 \left(4 c^4-c^2 (R+7)-1\right)}{c^2 \left(c^2-2\right)^2}\\
    {\mathfrak S}_1^4(R)&=\frac{s_0}{1-c^2},~~ ~~~{\mathfrak S}_1^5(R)=\frac{2s_0}{c^2-1}\\
     {\mathfrak S}_1^6(R)&={\mathfrak S}_1^7(R)=0\\
\end{split}
\end{equation}
(We have used the equation of motion to solve for $S_1$ in terms of the other scalar data. That is why ${\mathfrak S}_1^1(R)=0$).
\newline
Equation \eqref{escalar} is trivially solved by integration and we find
$$K_1^A(R)= -\int_{0}^Rd\rho~\mathfrak{S}_1^A(\rho).$$
We have chosen the integration constant above to satisfy the arbitrary 
condition $K_1^A(R=0)=0$. A different choice of integration constant could be 
absorbed by a shift of $\psi$ by a function of order $\frac{1}{d}$ and so 
is physically irrelevant. 

\subsection{Vector sector}
Here the relevant equations are $E_{Ry^i},~~E_{Vy^i}$ and $E_{Yy^i}$. We first 
study the linear combination of  $E_{Ry^i}$ and $E_{Vy^i}$  which yields 
a decoupled equation for $Q^A_2(R)$.
\begin{equation}\label{vectoreqn}
\begin{split}
E_i \equiv& ~2\left(e^{h_0} s_0^2 E_{Vy^i} + s_0 F(R) E_{R y^i}\right)\\
=&\sum_{A=1}^5 S_A\bigg(c^2F(R)\partial_R\left(e^{-R}Q^A_2(R)\right) + {\mathfrak V}^A_2(R)\bigg)=0
\end{split}
\end{equation}
where
\begin{equation}\label{vectorsource}
\begin{split}
{\mathfrak V}^1_2(R)&=-s_0(2-c^2) e^{2h_0-R}\\
{\mathfrak V}^2_2(R)&=s_0(1-c^2) e^{2h_0-R}\\
{\mathfrak V}^3_2(R)&=0,~~
{\mathfrak V}^4_2(R)=0,~~
{\mathfrak V}^5_2(R)=0\\
\end{split}
\end{equation}
Solving this equation with the boundary condition of `normalizability' 
at infinity we find 
\begin{equation}\label{q2samadhan}
Q_2^A(R) = \frac{e^R}{c^2}\int_R^\infty d\rho \left(\frac{{\mathfrak V}_2^A(\rho)}{F(\rho)}\right)
\end{equation}
Clearly $Q_2^A(R)$ above has a logarithmic singularity at $R=R_0$; the 
coefficient of this singularity is proportional to the vector equation of 
motion 
\begin{equation}\label{vectoreqn}
\text{Equation of Motion} \equiv \sum_{A=1}^5 V^\mu_A{\mathfrak V}_2^A(R_0)=0
\end{equation}
Upon substitution, \eqref{vectoreqn} reduces to the second of 
\eqref{mainequations}.  
Note, however, that all the nonzero source functions ${\mathfrak V}^A_2(R)$
are proportional to each other as functions of $R$
 (they are all proportional to 
$e^{-R}$). It follows that \eqref{vectoreqn} implies that 
$$\sum_{A=1}^5 V^\mu_A{\mathfrak V}_2^A(R)=0$$
at all $R$ and so it follows from  \eqref{q2samadhan} that 
$$Q_2(R) =0$$
Upon substituting the solution for $Q_2(R)$ and also the equation of motion, 
the equation  $E_{Ry^i}$ allows us to solve for $Q_1(R)$. It turns out that 
only $Q_1^1(R)$ has a nonzero source and we find
\begin{equation}\label{eqnq1}
\begin{split}
\frac{e^{h_0-R}}{4s_0}\partial_R\bigg(e^R\partial_R\left(e^{-R}Q_1^1(R)\right)\bigg) +\frac{e^{2h_0-R}}{2(1-c^2)}=0
\end{split}
\end{equation}
\eqref{eqnq1} is easily solved by two integrations. 
The solution is automatically regular everywhere. There are two integration 
constants. One of these is fixed by the requirement of `normalizability'
at $R=\infty$. The other integration constant can be absorbed into a 
shift in velocity field $u^\mu$ by a term of order $\frac{1}{d}$. We have 
arbitrarily fixed it by demanding that $Q_1(0)=0$.

\subsection{Tensor sector}
The equations are simplest in the symmetric tensor sector. 
The unique tensor equation, $E_{y^i y^j}$, evaluates to 
\begin{equation}\label{tensoreqn}
\begin{split}
E^{y^i y^j}\equiv -\sum_{A=1}^2[T_A]^{y^i y^j}e^{-R}\partial_R\bigg(e^RF(R)\partial_R\left(e^{-R}{\cal T}_A(R)\right)\bigg)=0
\end{split}
\end{equation}
Quite remarkably no tensor source appears on the RHS of \eqref{tensoreqn}. 
The most general solution of this equation is easily determined, and turns out 
to be either non-normalizable or singular at $R=R_0$. It follows that 
$${\cal T}_A(R)=0$$
and there are no corrections to the metric in the tensor sector.

\bibliography{mempap}
\end{document}